\documentclass[journal]{IEEEtran}

\usepackage{graphicx}
\usepackage{amsmath}
\usepackage{amssymb}
\usepackage{epstopdf}
\usepackage{epsf}
\usepackage{fancyhdr}
\usepackage{hyperref}

\usepackage[usenames]{color}
\usepackage{graphicx}
\usepackage{balance}
\usepackage{amsmath}
\usepackage{amssymb}
\usepackage{multirow}
\usepackage{cite}
\usepackage{ifpdf}
\usepackage{epstopdf}
\usepackage{epsf}
\usepackage{hyperref}
\usepackage{fancyhdr}

\begin{document}
\hyphenation{multi-symbol}
\abovedisplayskip=0.4pt
\belowdisplayskip=0.4pt
%\title{Parametric Field Estimation in Wireless Sensor Networks: Algorithms and Performance Evaluation}
\title{Distributed Estimation of a Parametric Field: \\ Algorithms and Performance Analysis}

\author{Salvatore~Talarico,~\IEEEmembership{Student Member,~IEEE,} Natalia~A.~Schmid,~\IEEEmembership{Member,~IEEE,} Marwan~Alkhweldi, and~Matthew~C.~Valenti,~\IEEEmembership{Senior~Member,~IEEE}% <-this % stops a space
\vspace{-0.4cm}
\thanks{Copyright (c) 2013 IEEE. Personal use of this material is permitted.
However, permission to use this material for any other purposes must be
obtained from the IEEE by sending a request to pubs-permissions@ieee.org. The
associate editor coordinating the review of this manuscript and approving it for
publication was Dr. Ruixin Niu. This work was supported by the Office of Naval Research
under Award No. N00014–09–1–1189. Portions of this paper were accepted for publication at the IEEE Military Communications Conference (MILCOM),Orlando,
FL, Oct. 2012.  }
\thanks{The  authors  are  with  the  Department  of  Computer  Science  and  Electrical Engineering, West Virginia University, Morgantown, WV 26506 USA
(e-mail:    salvatore.talarico81@gmail.com;    Natalia.Schmid@mail.wvu.edu;
malkhwel@mix.wvu.edu. Matthew.Valenti@mail.wvu.edu).}
\thanks{Color versions of one or more of the figures in this paper are available online
at http://ieeexplore.ieee.org.}
\thanks{Digital Object Identifier 10.1109/TSP.2013.2288684}
\thanks{}}
\date{}

%\pagestyle{fancy}
%%%%%%%%%%%%%%%%%%%%%%%%%%%%% The paper headers
%\fancyhead[RO,LE]{\small\thepage}
%\fancyhead[LO]{\small IEEE TRANSACTIONS ON SIGNAL PROCESSING, ACCEPTED FOR PUBLICATION}% odd page header and number to right top
%\fancyhead[RE]{\small TALARICO et al.: DISTRIBUTED  ESTIMATION OF A PARAMETRIC FIELD: ALGORITHMS AND PERFORMANCE ANALYSIS}%Even page header and number at left top
%\fancyfoot[L,R,C]{}
%\renewcommand{\headrulewidth}{0pt}% disable the underline of the header part

\maketitle

\thispagestyle{empty}
\vspace{-0.80cm}

\begin{abstract}
This paper presents a distributed estimator for a deterministic parametric physical field sensed by a homogeneous sensor network and develops a new transformed expression for the Cramer-Rao lower bound (CRLB) on the variance of distributed estimates. The proposed transformation reduces a multidimensional integral representation of the CRLB to an expression involving an infinite sum.  Stochastic models used in this paper assume additive noise in both the observation and transmission channels. Two cases of data transmission are considered. The first case assumes a linear analog modulation of raw observations prior to their transmission to a fusion center. In the second case, each sensor quantizes its observation to $M$ levels, and the quantized data are communicated to a fusion center. In both cases, parallel additive white Gaussian channels are assumed.  The paper develops an iterative expectation-maximization (EM) algorithm to estimate unknown parameters of a parametric field, and its linearized version is adopted for numerical analysis. The performance of the developed numerical solution is compared to the performance of a simple iterative approach based on Newton's approximation. While the developed solution has a higher complexity than Newton's solution, it is more robust with respect to the choice of initial parameters and has a better estimation accuracy.  Numerical examples are provided for the case of a field modeled as a Gaussian bell, and illustrate the advantages of using the transformed expression for the CRLB.  However, the distributed estimator and the derived CRLB are general and can be applied to any parametric field. The dependence of the mean-square error (MSE) on the number of quantization levels, the number of sensors in the network and the SNR of the observation and transmission channels are analyzed. The variance of the estimates is compared to the derived CRLB.
%, and the dependence on the density of the sensor network is considered.

\end{abstract}

\section{Introduction}
\label{sec:intro}

\PARstart{M}{any} military and civilian applications use distributed sensor networks as a platform to perform environmental monitoring, surveillance, detection, tracking, object classification and counting.
Typically, such applications impose constraints on power, bandwidth, latency, and/or complexity. Numerous research findings  have been reported on these topics over the past two decades (for example, see the reviews in \cite{WSN_survey_2002,WSN_optimization_2012}). Some publications are concerned with a specific application \cite{Bokareva2006,Yan2009,Ko10,Corke10,Wei11,Dietmar11,Lu2011}; other papers have formulated and solved problems in the area of distributed estimation, detection and tracking \cite{Patwari03,Liu04,Niu06,Jiang:2005,Chamberland07,levy08,Liu2009}. In particular, an important research thrust in the field of distributed estimation is the design of practical distributed algorithms, choosing either to optimize a distributed sensor network with respect to energy consumption during transmission \cite{Aravinthan06,Wu08,Li09} or to impose bandwidth constraints and thus focus on designing an optimal quantization strategy for the distributed network \cite{Xiao05, Ribeiro06, Niu2006, Aysal08}. Alternatively, both constraints could be considered \cite{Swami2007,Cui07}. For instance, Wu et al. \cite{Wu08} studied the problem of minimizing the estimation error, mean square error (MSE), under the constraint of limited power; based on the work in \cite{Wu08}, Li and Al-Regib  \cite{Li09} developed an upper bound on the lifetime of a wireless sensor network and then proposed a methodology to increase it; Xian and Luo \cite{Xiao05} provided a distributed estimation scheme that deals with low bandwidth channels efficiently; and Cui et al. \cite{Cui07}, imposing constraints on both power and bandwidth, proposed a transmission scheme that results in considerable power savings.

Among research groups working on the problem of distributed estimation, there are a few dealing with distributed estimation of a field (a multidimensional function, in general) \cite{Niu2006_1, Schabus11, Wang08, Nowak03, Schmid12}. In many real-world applications, distributed estimation of a multidimensional function may provide additional information that aids in making a high-fidelity decision or in solving another inference problem. Motivated by this, we contribute to this topic by formulating and solving the problem of a parametric field estimation from sparse noisy sensor measurements using an iterative solution. We also focus on the development of theoretical limits for distributed estimation of the parameters associated with a parametric field. These limits are used to compare the performance of the estimates obtained to the best performance that can be achieved in theory.

\begin{figure*}[ht]
\centering
 \vspace{-0.15cm}
\includegraphics[height=6cm,width=14cm]{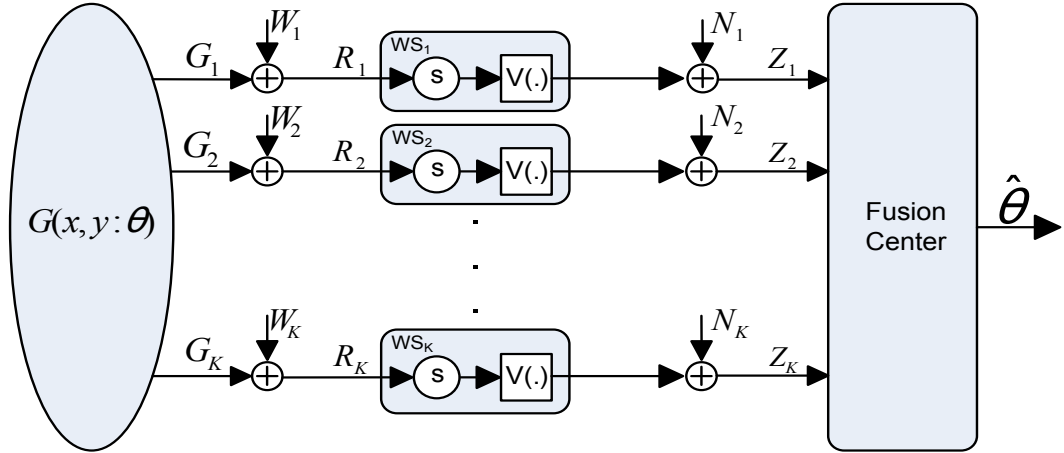}
 \vspace{0.05cm}
\caption{Block-diagram of the distributed sensor network comprised of K wireless sensors, denoted by $\mathsf{WS}_k,$ $k=1,\dots,K.$ } \label{fig:block_diagram}
 \vspace{-0.15cm}
\end{figure*}

In this paper, the problem of distributed estimation of a physical field from sensory data collected by a homogeneous sensor network is stated as a maximum-likelihood estimation problem. The physical field is a deterministic function and has a known spatial distribution parameterized by a set of unknown parameters, such as the location of an object generating the field or the strength and spread of the field in the region occupied by the sensors. It is assumed that white Gaussian noise is added at the sensors and over the transmission channel from the sensors to a fusion center, and that signal processing is performed both locally at the sensor nodes as well as globally at the fusion center.

Two cases of transmission channels are considered. The first case, which we refer to as a \textit{amplify-and-forward}, assumes that noisy sensor measurements are transmitted as analog signals to a fusion center. The second case, which we refer to as a \textit{quantize-and-forward}, assumes that each sensor locally quantizes its observation to $M$ levels, and the quantized data are digitally modulated and communicated to a fusion center. For both cases, orthogonal additive white Gaussian noise channels are considered and the effect of interference is neglected, since we assume an interference avoidance protocol is used.

An iterative algorithm to estimate unknown parameters is formulated, and a simplified numerical solution involving additional approximations is developed for both cases of the transmission channel. Furthermore, a new transformed expression for the CRLB on the variance of distributed estimates of field parameters is derived. The applied transformation reduces a multidimensional integral to a single infinite series.  Although it is an infinite series, only the first few terms need to be evaluated in order to obtain a reasonable accuracy.   The developed CRLB is applied to evaluate the performance of the estimates of the field parameters for both types of transmission channels.  The developed expectation-maximization (EM) solution and the bound are general. However, this work uses a Gaussian bell function as the parametric field to illustrate the EM approach and the derived bound. Our numerical evaluation shows that the variance of the EM estimates approaches the CRLB as the density of the network increases, provided initial values of the estimates are selected sufficiently close to the true parameters.  The accuracy, performance, and the rate of convergence of the developed EM algorithm are compared to those of the Newton-Raphson (NR) algorithm.  Numerical results highlight the advantages and disadvantages of the developed EM algorithm.

The rest of the paper is organized as follows. Sec. \ref{sec:Network Model} describes the network. Sec. \ref{sec:Distributed parameter estimator} describes the proposed EM solution. Sec. \ref{sec:Cramer_Rao_Lower_Bound} develops the CRLB on the variance of parameter estimates. Sec. \ref{sec:Numerical_Analysis} provides numerical performance evaluation of the estimator and compares the variance of the estimated parameters to the CRLB. Finally, Sec. \ref{sec:Summary} presents a summary of the results.

\section{Network Model}
\label{sec:Network Model}
Consider a distributed network of homogeneous sensors monitoring the environment for the presence of a substance or an object. Assume that each substance or object is characterized by one or more location parameters and by a spatially distributed physical field generated by them.
Examples of physical fields include (1) a magnetic field generated by a dipole and modeled as a function of the inverse cube of the distance to the dipole \cite{Caruso2000,Marshall1978,Bapat2005}, (2) a radioactive field modeled as a stationary spatially distributed Poisson field with a two-dimensional intensity function decaying according to the inverse-square law \cite{Snyder1991} or (3) a cloud of pollution or chemical fumes \cite{Nehorai95} that, if stationary, has a spatial intensity that can often be modeled as a Gaussian bell.

A block diagram of a distributed sensor network used for the estimation of parameters of a physical field is shown in Fig. \ref{fig:block_diagram}. The network is composed of $K$ sensors randomly distributed over a finite area $A$. The network is calibrated, and the relative locations of the sensors are known, which are common assumptions \cite{Jiang:2005,Niu2006}. Sensors act independently of one another and take noisy measurements of a physical field $G(x,y)$ defined over the area $A.$  A sample of $G(x,y)$ at a location $(x_k,y_k)$  is denoted as $G_k=G(x_k,y_k),$ $k=1,\dots,K$. The parametric field $G(x,y)$ is characterized by $L$ unknown parameters $\mathbf{\boldsymbol \theta}=[\theta_1,\ldots,\theta_L]^T.$  We use $G(x_k,y_k:\boldsymbol \theta)$ to emphasize this dependence and use $G_k$ for brevity.   The sensor noise at different locations, denoted by $W_k,$ $k=1,\ldots,K$, is modeled as independent Gaussian random variables with zero mean and known variance $\sigma^2_k$.  Let $R_k,$ $k=1,\ldots,K$, be the noisy samples of the field at the location of distributed sensors. Then  $R_k$ is modeled as $R_k=G_k+W_k.$  These observations are transmitted over noisy parallel channels to a {\em fusion center} (FC), which estimates the vector parameter $\mathbf{\boldsymbol \theta}.$   We assume that the communication channels do not interfere. The method used to send these observations is chosen accordingly to the application constraints and the channel characteristics, and it is denoted in Fig.  \ref{fig:block_diagram} by $\mathcal V(\cdot)$.

If the sensor observations are transmitted as analog samples over the channels without any prior processing (aside from linearly modulating with a suitable carrier frequency $f_c$ and an amplification to assure the transmit power is at a desired level), we refer to it as a \textit{amplify-and-forward}. Denote by $Z_1,\ldots,Z_K$ the noisy observations received by the FC and obtained by coherently demodulating the signal sent by the sensors, as shown in Fig. \ref{fig:block_diagram}. In this case, the observations $Z_k$ are given as $Z_k = R_k + N_k$, $k = 1,\dots,K$, where the noise $N_k$ in the channel $k$ is white Gaussian with variance $\eta_k^2$, and $R_k$ and $N_k$ are independent \footnote{The channel gain is normalized to unity and the effect of fading is absorbed into the variance of the transmission channel noise.}.

Due to constraints imposed by practical technology, each sensor may be required to quantize its measurements prior to transmitting them to the FC.  Assume a deterministic quantizer with $M$ quantization levels at each sensor location.  Let  $\{\nu_1,\nu_2,\ldots,\nu_M\}$  be the {\em reproduction points} of a quantizer and $\tau_1=-\infty,\tau_2,\ldots,\tau_{M},\tau_{M+1}=\infty$ be the boundaries of the quantization regions.  Then the output of the $k$-th quantizer is a random variable $q(R_k)=q_k$ taking value $\nu_j$ with probability
%\vspace{-0.3cm}
\begin{eqnarray}
 p_{k,j}(\boldsymbol \theta ) &=& \int_{\tau_j}^{\tau_{j+1}} \frac{1}{\sqrt{2\pi\sigma^2_k} } \exp \left( -\frac{ \left( t -G_k \right)^2 }{2\sigma^2_k } \right)dt,
\label{eq:p_kj}
\end{eqnarray}
which depends on the unknown parameters $\boldsymbol \theta$ of the physical field $G_k$.

The quantized data are modulated using a linear digital modulation scheme, such as on-off keying (OOK), an M-ary quadrature amplitude modulation (QAM) or a pulse amplitude modulation (PAM), and then transmitted to the FC over noisy parallel channels. We refer to this channel as a \textit{quantize-and-forward}.  To further clarify the modulation and demodulation steps, denote by  $\mathbf{B}_k=[B_k(1),\ldots ,B_k(\log_2(M))]^T$ a bit representation of the quantized observation of the $k$-th sensor in vector form.  Then by applying a linear modulation scheme (OOK or QAM, for example) and then coherently demodulating it, the vector of observed data at the FC will be in the form $\mathbf{Z}_k=\mathbf{B}_k+\mathbf{N}_k,$ where $\mathbf{N}_k$ is a white Gaussian noise vector independent of $\mathbf{B}_k.$
% Note that in both cases, the FC coherently demodulates the observed data.
%The noise of the digital channel is due to the channel impairments, and is denoted by $\tilde{N}_k,$ $k=1,\dots,K.$ If $m(\cdot)$ represents a modulation function and $d(\cdot)$ represents a demodulation function, then each $Z_k$ is given by $Z_k=d\left(m(q_k)+\tilde{N}_k\right), \ \  k=1,\dots,K.$ After demodulation and in particular the correlator, $Z_k$ can be written as $Z_k=q_k+N_k,$ where $N_k$ is again white Gaussian noise with variance $\eta^2.$ For both analog and digital channel, since we are not interested in estimating the transmit signal, but ultimately in the estimation of a parametric field, the knowledge of the distribution of $Z_k$ is used in the following sections without any need to estimate the received signal by applying a maximum a posteriori probability (MAP) criterion.
\section{Distributed parameter estimator}
\label{sec:Distributed parameter estimator}
Given the noisy measurements and the relative location of the sensors in the network, the task of the FC is to estimate the vector parameter $\mathbf{\boldsymbol \theta}.$ In this work, the maximum-likelihood (ML) estimation approach \cite{Trees2001, Poor1994} is used to solve the problem of distributed parameter estimation for both channels. Denote by $l(\mathbf{Z}:\boldsymbol \theta)$ the log-likelihood function of the vector of the random measurements $\mathbf{Z}$ parameterized by vector $\boldsymbol \theta$, where $\mathbf{Z}$ is the vector of measurements $[Z_1,Z_2,\ldots,Z_K]^T.$ To be more specific,  $l(\mathbf{Z}:\boldsymbol \theta)$ is defined as $\log f_{\mathbf{Z}:\boldsymbol \theta},$ with $f_{\mathbf{Z}:\boldsymbol \theta}$ being the parameterized probability density function (pdf) of the vector $\mathbf{Z}.$ The ML solution is the vector $\boldsymbol \theta$ that maximizes the log-likelihood function $l(\mathbf{Z}:\boldsymbol \theta)$:
%\vspace{-0.1cm}
\begin{eqnarray}
 \hat{\mathbf{\boldsymbol \theta}}& = &\mbox{arg}\max_{\mathbf{\boldsymbol \theta}\in {\boldsymbol \Theta}} l(\mathbf{Z}:\mathbf{\boldsymbol \theta}),
\label{eq:incomplete_likelihood1}
\end{eqnarray}
where $\boldsymbol \Theta$ is a set of admissible vector parameters.\\
The necessary conditions to find the maximizer are given by:
%\vspace{-0.1cm}
\begin{eqnarray}
\nabla_{\boldsymbol \theta}l(\mathbf{Z}: \boldsymbol \theta)\Big|_{\hat{\boldsymbol \theta }}=0,
\label{eq:likelihood_equation}
\end{eqnarray}
where $\nabla_{\boldsymbol \theta}$ denotes the gradient with respect to the vector $\boldsymbol \theta,$ and the maximizer is the interior point of $\boldsymbol \Theta.$

\subsection{Amplify-and-forward channel}
\label{EAC}
For the amplify-and-forward channel, the signals received at the FC are independent but not identically distributed and the log-likelihood function is given as:
%\vspace{-0.2cm}
\begin{eqnarray}
\hspace{-0.25 cm} l(\mathbf{Z}:\boldsymbol \theta) \hspace{-0.25 cm} &=& \hspace{-0.25 cm} \sum^{K}_{k =1} \log(f_{Z_{k}}(Z_k))\nonumber \\
&=& \hspace{-0.25 cm} -\frac{1}{2} \sum^{K}_{k =1} \frac{\left[ Z_k - G\left( x_k,y_k:\mathbf{\boldsymbol \theta}\right) \right]^2}{(\sigma^2_k+\eta^2_k)} + C_1,
\label{eq:complete_log}
\end{eqnarray}
where $C_1$ is not a function of $\mathbf{\mathbf{\boldsymbol \theta}}$.

The condition (\ref{eq:likelihood_equation}) applied to (\ref{eq:complete_log}) generates a set of nonlinear equations in ${\boldsymbol \theta}.$
%To simplify the solution, we linearize the expression in (\ref{eq:complete_log})
The solution to (\ref{eq:incomplete_likelihood1}) is found numerically by means of Newton's method \cite{chang2005}.  This solution is applied to generate the results in Sec. \ref{numerical_analysis_analog_case}.
\vspace{-0.1cm}
\subsection{Quantize-and-forward channel}
\label{EDC}
For the quantize-and-forward channel, the joint likelihood function of the independent quantized noisy measurements $\mathbf{Z}_1, \mathbf{Z}_2, \ldots, \mathbf{Z}_K$ can be written as
\small
\begin{eqnarray}
\hspace{-0.15 cm} l(\mathbf{Z}:\boldsymbol \theta) \hspace{-0.3 cm} & = & \hspace{-0.4 cm} \sum^{K}_{k =1} \log \hspace{-0.05 cm} \left[\sum^{M}_{j =1}\hspace{-0.2 cm} \ p_{k,j}(\boldsymbol \theta )  \exp \hspace{-0.05 cm} \left( \hspace{-0.15 cm} - \frac{ \left( \mathbf{Z}_k - \mathbf{b}_j \right)^T \left( \mathbf{Z}_k - \mathbf{b}_j \right) }{2\eta^2_k } \hspace{-0.10 cm}\right)\hspace{-0.1 cm}\right] \hspace{-0.2 cm} + \hspace{-0.1 cm} C_2,
\label{eq:incomplete_likelihood}  \hspace{0.4 cm}
\end{eqnarray}
\normalsize
where $p_{k,j}(\boldsymbol \theta)$ is defined in (\ref{eq:p_kj}), $\mathbf{b}_j$ is a binary representation of the integer $j$ and, thus, of $\nu_j$, characterized by $\alpha=\log_2\left(M\right)$ bits; $C_2$ is not a function of $\mathbf{\mathbf{\boldsymbol \theta}}$.
%%$C_2 = -\frac{K}{2} \log\left( 2 \pi \eta^2\right)$
%and $p_{k,j}$ is the probability that sensor $k$ quantizes its noisy observation into the $j$-th reproduction point, and is
%given by
%\[ p_{k,j}(\boldsymbol \theta ) =\int_{\tau_j}^{\tau_{j+1}} \frac{1}{\sqrt{2\pi\sigma^2_k} } \exp \left( -\frac{ \left( t -G_k \right)^2 }{2\sigma^2_k } \right)dt.\]
The ML solution $\mathbf{\hat{\boldsymbol \theta}}$ is the solution that maximizes the expression (\ref{eq:incomplete_likelihood}), but since applying (\ref{eq:likelihood_equation}) to (\ref{eq:incomplete_likelihood}) results in a set of nonlinear equations, an iterative solution to the problem can be developed: (1) a set of EM iterations \cite{Dempster77} are formulated and then (2) a Newton's linearization is used to solve EM equations for the unknown parameters.

\subsubsection{Expectation Maximization Solution}
The random variables $(R_k,\mathbf{N}_k)$, $k=1,2, \dots, K$ are selected as complete data. The complete data log-likelihood, denoted by $l_{cd}(\cdot)$, is given by
%\vspace{-0.15cm}
\begin{equation}
l_{cd}(\mathbf{R},\mathbf{N})=  - \frac{ 1 }{2 }\sum_{k=1}^{K}\frac{(R_k-G_k)^2}{\sigma^2_k} +C_3,
\end{equation}
where $C_3$ is not a function of $\mathbf{\boldsymbol \theta}.$
The measurements $\mathbf{Z}_k,$ $k=1,\dots,K,$ form incomplete data and the mapping from complete data space to incomplete data space is given by $\mathbf{Z}_k=\mathbf{B}_k+\mathbf{N}_k.$

Denote by $\hat{\mathbf{\boldsymbol \theta}}^{(m)}$ an estimate of the vector ${\boldsymbol \theta}$ obtained at the $m$-th iteration.  To update the estimates, the expectation and maximization steps are alternated. During the expectation step, the conditional expectation of the complete data log-likelihood is evaluated as follows:
%\vspace{-0.15cm}
\begin{eqnarray}
T^{(m+1)}& = &E\left[ \left. - \frac{ 1 }{2 }\sum^{K}_{k =1}\frac{(R_k-G_k)^2}{\sigma^2_k} \right| {\mathbf{Z},\hat{\boldsymbol \theta }^{(m)}}\right],
\label{eq:E-step}
\end{eqnarray}
where the expectation is with respect to complete data, given the incomplete data (measurements) and the estimates of the parameters at the $m$-th iteration. During the maximization step, (\ref{eq:E-step}) is maximized:
%\vspace{-0.15cm}
\begin{multline}
\hspace{-0.2cm} \frac{dT^{(m+1)}}{d\theta_t} = \left. E\left[ \left. \sum^{K}_{k =1}\frac{(R_k-G_k)}{\sigma^2_k}\left(\frac{dG_k}{d\theta_t}\right) \right| {\mathbf{Z},\hat{\mathbf{\boldsymbol \theta} }^{(m)}}\right] \right|_{\hat{\boldsymbol \theta}^{(m+1)}} \hspace{-0.5cm}=0,\\\text{ $ t$}=1,\ldots,L.
\label{eq:M-step}
\end{multline}
To find the expectation, it can be noted that the conditional probability density function of $\mathbf{Z}_k,$  given $R_k,$  is Gaussian with mean $\mathbf{B}_k$  and a diagonal covariance matrix with identical diagonal entries $\eta^2_k$  and the pdf of $R_k$ is Gaussian with mean $G_k$  and variance $\sigma^2_k.$  It can be also noted that at the $m$-th iteration the conditional pdf of $R_k,$   given $\mathbf{Z}_k,$   implicitly involves the estimates of the parameters obtained at the $m$-th iteration.

Denote by $G_k^{(m)}$ the estimate of the field $G(x,y)$ at the location $(x_k,y_k)$ with the vector of parameters ${\boldsymbol \theta}$ replaced by their estimates $\hat{\mathbf{\boldsymbol \theta}}^{(m)}.$ Then the following lemma can be formulated.

\newtheorem{theorem}{Lemma 3. \hspace{-0.25 cm}}
\begin{theorem}
The expression for the iterative evaluation of the unknown parameters can be written as:
%\vspace{-0.25cm}
\begin{multline}
\hspace{-0.3cm} \sum^{K}_{k =1}\frac{dG_k^{(m+1)}}{d\theta_t}A(G_k^{(m)})-\sum^{K}_{k =1}G_k^{(m+1)}\frac{dG_k^{(m+1)}}{d\theta_t}B(G_k^{(m)})
=0, \\ t=1,\ldots,L,
\label{eq:structure}
\end{multline}
%\vspace{-0.2cm}
where
%\vspace{-0.15cm}
\begin{eqnarray}
A(G_k^{(m)})\hspace{-0.3 cm}&=&\hspace{-0.3 cm}\sum_{j=1}^M \hspace{-0.1 cm} \frac{\exp\left(-\frac{(\mathbf{z}_k-\mathbf{b}_j)^T(\mathbf{z}_k-\mathbf{b}_j)}{2\eta^2_k}\right)}
{f_{\mathbf{Z}_k}^{(m)}(\mathbf{z}_k)\left( 2 \pi \eta^2_k\right)^{\alpha/2}} \hspace{-0.1 cm} \left( \hspace{-0.1 cm} \sqrt{\frac{\sigma^2_k}{2\pi}} e^{-\frac{(\tau_j-G_k^{(m)})^2}{2\sigma^2_k}} \right. \nonumber \\
& & \hspace{-0.6 cm} \left. - \sqrt{\frac{\sigma^2_k}{2\pi}} e^{-\frac{(\tau_{j+1}-G_k^{(m)})^2}{2\sigma^2_k}} + G_k^{(m)} \Delta T^{(m)}(j,k)\right),
\end{eqnarray}
%\vspace{-0.4cm}
\begin{eqnarray}
B(G_k^{(m)})=\sum^{M}_{j =1}\frac{ \exp \left(-\frac{(\mathbf{z}_k-\mathbf{b}_j)^T(\mathbf{z}_k-\mathbf{b}_j) }{2\eta^2_k }\right)}{f_{\mathbf{Z}_k}^{(m)}(\mathbf{z}_k) \left( 2 \pi \eta^2_k\right)^{\alpha/2}}\Delta T^{(m)}(j,k),
\end{eqnarray}
with
%\vspace{-0.2cm}
\begin{equation}
\Delta T^{(m)}(j,k)= Q \hspace{-0.05 cm} \left(\frac{\tau_j-G_k^{(m)}}{\sigma_k}\right)-Q \hspace{-0.05 cm} \left(\frac{\tau_{j+1}-G_k^{(m)}}{\sigma_k}\right),
\end{equation}
%\vspace{-0.2cm}
\begin{equation}
f_{\mathbf{Z}_k}^{(m)}(\mathbf{z}_k)=\int f^{(m)}_{\mathbf{Z}_k|R}(\mathbf{z}_k|r) f^{(m)}_{R_k}(r) dr.
\end{equation}
The expression $Q(\cdot)$ is used to denote the Q-function, that is given by
%\vspace{-0.4cm}
\begin{equation}
   Q(x)=\frac{1}{\sqrt{2\pi}}\int_{x}^{\infty} \exp \left( -\frac{t^2}{2} \right ) dt.
\end{equation}
\end{theorem}
Proof: The details of the derivation are moved to the Appendix A.

\subsubsection{Linearization}
\label{sec:Linearization}
The equations (\ref{eq:structure}) are nonlinear in $\hat{\mathbf{\boldsymbol \theta}}^{(m+1)}$ and have to be solved numerically for each iteration. To simplify the solution, the expression in (\ref{eq:structure}) is linearized  by means of Newton's method. Denote by $\mathbf{F}(\boldsymbol \theta^{(m+1)})$ the vector form of the first derivative of the left side in (\ref{eq:structure}) over $\boldsymbol \theta^{(m+1)}$ and let $J\left(\boldsymbol \theta_n^{(m+1)}\right)$ be the Jacobian of the left side in (\ref{eq:structure}) . The index $n$ indicates the iteration of the Newton's solution. Then $\mathbf{\boldsymbol \theta}^{(m+1)}$ solves the following linearized equation:
%\vspace{-0.1cm}
\begin{eqnarray} \label{}
J\left(\boldsymbol \theta_n^{(m+1)}\right)\left[\boldsymbol \theta_{n+1}^{(m+1)}-\boldsymbol \theta_{n}^{(m+1)}\right]=-F \left( \boldsymbol \theta_n^{(m+1)}\right).
\end{eqnarray}

\subsubsection{Alternative solution}
\label{sec:Alternative solution}
For a parametric field estimation with the network model described in Sec. \ref{sec:Network Model} the Newton-Raphson (NR) algorithm \cite{Kobayashi:2012} can be used as an alternative to the EM algorithm.  The EM and NR algorithms have a similar computational complexity that becomes imperceptible as the network becomes sparse and/or as the number of quantization levels decreases. The NR algorithm displays a faster convergence compared to the EM algorithm.  However, the proposed EM solution converges more consistently to the local maximum of the likelihood function \cite{Lindstrom1988} and guarantees a better accuracy. Sec. \ref{sec:Comparison Between EM and NN} provides an illustration, where the EM and NR algorithms are compared both in terms of rate and consistency of convergence as well as in terms of precision of the final results.

\vspace{-0.2cm}
\section{Cramer-Rao Lower Bound}
\label{sec:Cramer_Rao_Lower_Bound}
In Sec.  \ref{sec:Distributed parameter estimator}, parameters of the physical field $G(x,y)$ were estimated for the amplify-and-forward and the quantize-and-forward channel by means of a
Newton's method and a linearized version of EM algorithm, respectively. In order to evaluate the efficiency of the estimator for both channels the CRLB on the variances of unknown parameters is evaluated.

In this section, we assume that the parameter estimates are unbiased. Denoted by $\Sigma_{\large \boldsymbol \theta}$ the covariance matrix of the estimated parameters $\hat{\boldsymbol \theta},$ the CRLB on $\Sigma_{\large \boldsymbol \theta}$ \cite{Trees2001, Poor1994} is given as
%\vspace{-0.3cm}
\begin{eqnarray}
\Sigma_{\large \boldsymbol \theta} \geq I^{-1}(\boldsymbol \theta), \label{eq3}
\end{eqnarray}
where $I(\boldsymbol \theta)$ denotes the Fisher information matrix with the entry at the location $(s,t)$ given by
%\vspace{-0.1cm}
\begin{eqnarray}
I_{s,t} &=& -E \left[ \frac{\partial^2 l(\mathbf{Z}:\boldsymbol \theta) }{\partial \theta_s \partial \theta_t }\right]. \label{Fisher_Information_2}
\end{eqnarray}
Below we detail the derivation of the CRLB for the amplify-and-forward and the quantize-and-forward channel.
\vspace{-0.4cm}
\subsection{Amplify-and-forward channel}
 In the case of amplify-and-forward channel, substituting (\ref{eq:complete_log}) into (\ref{Fisher_Information_2}), and simplifying all the terms that do not depend on $\boldsymbol \theta$, yields
 %\vspace{-0.2cm}
\begin{eqnarray}
  I_{s,t}  =\hspace{-0.1cm}  \frac{1}{2}\hspace{-0.1cm} \sum_{ k=1}^{ K} \hspace{-0.05cm} \frac{1}{(\sigma^2_k + \eta^2_k )}E\left[ \frac{\partial^2}{\partial \theta_s \partial \theta_t } \hspace{-0.05cm} \left( Z_k - G(x_k, y_k: \hspace{-0.05cm} \boldsymbol \theta)\right)^2 \right].
  \label{Analog_eq2}
\end{eqnarray}
After taking the partial derivatives and noting that
$E\left[ Z_k - G(x_k, y_k: \boldsymbol \theta)\right]=0$,  (\ref{Analog_eq2}) becomes
%\vspace{-0.2cm}
\begin{eqnarray}
  I_{s,t} = - \sum_{ k=1}^{ K} \frac{1}{(\sigma^2_k + \eta^2_k )}\frac{\partial G( x_k, y_k: \boldsymbol \theta)}{\partial \theta_s} \frac{\partial G( x_k, y_k: \boldsymbol \theta)}{\partial \theta_t}.
  \label{Analog_eq3}
\end{eqnarray}
\vspace{-0.6cm}
\subsection{Quantize-and-forward channel}
In the case of quantize-and-forward channel, substituting (\ref{eq:incomplete_likelihood}) into (\ref{Fisher_Information_2}) yields
%\vspace{-0.3cm}
\begin{eqnarray}
  I_{s,t} =  -E\left[ \frac{\partial^2 }{\partial \theta_s \partial \theta_t } \sum_{ k=1}^{ K} \log  x_k \left(\boldsymbol \theta \right)  \right],
  \label{Fisher_eq1}
\end{eqnarray}

\begin{figure*}[ht]
\vspace{-0.25cm}
\setcounter{equation}{26}
\begin{eqnarray}
\hspace{-0.5cm} I_{s,t} \hspace{-0.25 cm} &=& \hspace{-0.25 cm} - \sum_{ k=1}^{ K}  \sum_{j=1}^{M} \left \{ \frac{\partial^2  p_{k,j}\left( \boldsymbol \theta \right)}{\partial \theta_s \partial \theta_t} - \sum_{i=1}^{M}  \frac{\partial
p_{k,j}\left(\boldsymbol \theta \right) }{\partial \theta_s}  \frac{\partial p_{k,i}\left(\boldsymbol \theta \right) }{\partial \theta_t} \sum_{n=0}^{\infty } \sum_{m=0}^n \hspace{-0.15 cm}\left( -1\right)^{m+n} \frac{n!}{m!}
\hspace{-0.65cm} \mathop{ \sum_{\ell_v \geq 0}}_{\sum_{v=0}^{M}\ell_v=n-m} \hspace{-0.65cm}
   \left[ \left(\prod_{v=1}^M \frac{ p_{k,v}\left( \boldsymbol \theta \right) }{\ell_v !}\right)  \Lambda_{\ell_v} \right] \right\}\label{FinalExpression}
\end{eqnarray}
\hrulefill
\vspace{-0.55cm}
\end{figure*}
\setcounter{equation}{20}

where $x_k \left( \boldsymbol \theta \right)$ is given by the following expression
%\vspace{-0.2cm}
\begin{eqnarray}
x_k \left( \boldsymbol \theta \right) =  \sum_{j=1}^{M} p_{k,j}\left( \boldsymbol \theta \right) \exp \left( -\frac{(\mathbf{Z}_k-\mathbf{b}_j)^T(\mathbf{Z}_k-\mathbf{b}_j)}{2 \eta^2_k} \right).
  \label{Fisher_eq2}
\end{eqnarray}
After taking the partial derivatives in (\ref{Fisher_eq1}) we have
%\vspace{-0.2cm}
\small
\begin{eqnarray}
  I_{s,t} \hspace{-0.10cm}
  = - \sum_{ k=1}^{ K}  E \hspace{-0.1 cm}\left[\frac{1}{x_k \left( \boldsymbol \theta \right)} \frac{\partial^2 x_k \left( \boldsymbol \theta \right)}{\partial \theta_s \partial \theta_t} -  \frac{1}{x_k^2 \left( \boldsymbol \theta \right)} \frac{\partial x_k
  \left( \boldsymbol \theta \right)}{\partial \theta_t} \frac{\partial x_k \left( \boldsymbol \theta \right)}{\partial \theta_s} \right]\hspace{-0.1 cm}. \label{Fisher_eq3}
\end{eqnarray}
\normalsize
Substituting (\ref{Fisher_eq2}) into (\ref{Fisher_eq3}) yields
%\vspace{-0.15cm}
\small
\begin{eqnarray}
  \hspace{-0.1 cm} I_{s,t} \hspace{-0.1 cm}= \hspace{-0.1 cm}-   \hspace{-0.1 cm} \sum_{ k=1}^{ K}   \hspace{-0.05 cm} \sum_{j=1}^{M} \hspace{-0.1 cm} \left \{   \hspace{-0.1 cm}\frac{\partial^2  p_{k,j}\left( \boldsymbol \theta \right)}{\partial \theta_s \partial \theta_t} \Gamma_{k,j}   \hspace{-0.1 cm}-   \hspace{-0.1 cm} \sum_{i=1}^{M}
    \hspace{-0.1 cm} \frac{\partial p_{k,j}\left( \boldsymbol \theta \right) }{\partial \theta_s}  \frac{\partial p_{k,i}\left( \boldsymbol \theta \right) }{\partial \theta_t} \Phi_{k,j,i}  \hspace{-0.1 cm} \right\}  \hspace{-0.1 cm} \label{Fisher_eq4},   \hspace{-0.1 cm}
\end{eqnarray}
\normalsize where
%\vspace{-0.15cm}
\small
\begin{eqnarray}
\Gamma_{k,j}  =  E\left[ \frac{\exp \left( -\frac{(\mathbf{Z}_k-\mathbf{b}_j)^T(\mathbf{Z}_k-\mathbf{b}_j)}{2 \eta^2_k} \right) }{\sum_{v=1}^{M} p_{k,v}\left( \boldsymbol \theta \right) \exp \left( -\frac{(\mathbf{Z}_k-\mathbf{b}_v)^T(\mathbf{Z}_k-\mathbf{b}_v)}{2 \eta^2_k} \right)}\right],
\label{Fisher_eq5} \\ \Phi_{k,j,i}  = E\left[ \frac{\exp \left( -\frac{(\mathbf{Z}_k-\mathbf{b}_j)^T(\mathbf{Z}_k-\mathbf{b}_j)+(\mathbf{Z}_k-\mathbf{b}_i)^T(\mathbf{Z}_k-\mathbf{b}_i)}{2 \eta^2_k}\right)}{\left[ \sum_{v=1}^{M}
p_{k,v}\left( \boldsymbol \theta \right) \exp \left( -\frac{(\mathbf{Z}_k-\mathbf{b}_v)^T(\mathbf{Z}_k-\mathbf{b}_v)}{2 \eta^2_k} \right) \right]^2} \right]. \label{Fisher_eq6}
\end{eqnarray}
\normalsize
Since the expectation in (\ref{Fisher_eq5}) is with respect to the pdf of $Z_k,$ which is given by
%\vspace{-0.10cm}
\small
\begin{eqnarray}
f_{\mathbf{Z}_k}(\mathbf{z}_k) = \sum_{v=1}^{M}  \frac{p_{k,v}\left(\boldsymbol \theta \right)}{\left( 2 \pi \eta^2_k\right)^{\alpha/2}} \exp \left( -\frac{(\mathbf{z}_k-\mathbf{b}_v)^T(\mathbf{z}_k-\mathbf{b}_v)}{2 \eta^2_k} \right) \label{pdf_Z_k},
\end{eqnarray}
\normalsize
the expression (\ref{Fisher_eq5}) integrates to 1, that is,
$\Gamma_{k,j}=1.$

\newtheorem{theorem1}{Lemma 4. \hspace{-0.25 cm}}
\begin{theorem1} The $I_{s,t}$ entry of the Fisher information matrix (\ref{eq3}) can be transformed in (\ref{FinalExpression}) (displayed at the top of this page) where $\Lambda_{\ell_v}$ is given by (\ref{Fisher_eq12}).\setcounter{equation}{27}
\end{theorem1}

Proof: The details of the proof are presented in the Appendix B.

Thus the entry $I_{s,t}$ of the Fisher information matrix can be implemented using two approaches: (i) by evaluating numerically (\ref{Fisher_eq1}) and (ii) by truncating the infinite sum in (\ref{FinalExpression}) to $\zeta$ terms such that a compromise between computational complexity and accuracy is achieved. We involve Simpson's method \cite{Atkinson:1989} as an alternative approach to evaluate $I_{s,t}.$  It provides a good tradeoff between accuracy and speed.  The results from the two different implementations coincide even when very few terms $\zeta$ are used in the series representation (\ref{FinalExpression}).  This is demonstrated at the end of Sec. \ref{sec:Digital channel} that presents a comparison of the two implementations.

\vspace{-0.2 cm}
\section{Numerical Analysis}
\label{sec:Numerical_Analysis}
In this section, the performance of the distributed ML estimator is evaluated for both types of channels. Numerical results are presented for the case of the field modeled as a Gaussian bell.
%and by the CRLB, derived in Sec. \ref{sec:Cramer_Rao_Lower_Bound}, its efficiency will be checked, using a Gaussian shaped field as an example.
However, the distributed ML estimator in Sec.\ref{EAC} and Sec.\ref{EDC} and the bounds in (\ref{Analog_eq3}) and (\ref{FinalExpression}) are general and can be applied to estimates of any parametric field $G(x,y).$

Our numerical analysis assumes that a distributed network of $K$ sensors is formed by deploying them uniformly at random over a finite area $A$ of size $8 \times 8$, where the location of each sensor is noted. The Gaussian field used in our simulations is
%\vspace{-0.2 cm}
\begin{eqnarray}
G(x_k,y_k)= h \exp\left[ -\frac{\left( x_k-x_c \right)^2}{2 \rho_x^2} -\frac{\left( y_k-y_c \right)^2}{2 \rho_y^2} \right],
\end{eqnarray}
where $h$ is the ``strength'' of the field, $\rho_x^2$ and $\rho_y^2$ determine the ``spread'' of the bell in the x and y direction, respectively, and $\left(x_c,y_c\right)$ is the position of the object generating the field. In the numerical examples, we set $h=8$, $\rho_x^2=\rho_y^2=4.$ The location parameters of the field are set to $x_c=4$ and $y_c=4$.
For numerical illustration we assume that $h$, $\rho_x$, $\rho_y$, $x_c$ and $y_c$ are all unknown parameters that have to be estimated; i.e., in our experiments the unknown vector-parameter is $\boldsymbol \theta = [h, \rho_x, \rho_y, x_c, y_c]$.  The size of the network $K$ is varied from $10$ to $200$.
A Gaussian field is sampled at the location of the $k$-th sensor, $k=1,\ldots,K$, and a sample of randomly generated Gaussian noise with mean zero and variance  $\sigma^2$ is added to each field measurement.  Note that our experiments assume i.i.d. noise samples, and, for simplicity, $\sigma^2_k=\sigma^2$ and $\eta^2_k=\eta^2$ for all sensors.
%A Gaussian field shown in Fig. \ref{fig:Gaussian_field} is sampled at the location of the $k$-th sensor, $k=1,\ldots,K$ and a sample of randomly generated Gaussian noise with mean zero and variance $\sigma^2$ is added to each field measurement.
%\begin{figure}[!t]
%\centering
%\includegraphics[width=9cm]{figures/gaussian_field}
%\vspace{-0.75cm}
%\caption{Gaussian field located at $(x_c,y_c)=(4,4)$ with peak parameter $8$ and variance $4.$ }
%\label{fig:Gaussian_field}
%\vspace{-0.45cm}
%\end{figure}
The noise variance $\sigma^2$ is selected such that the total signal-to-noise ratio (SNR) of the local observations defined as
%\vspace{-0.2 cm}
\begin{equation}
\mathsf{SNR_O} = \frac{ {\int \int}_A G^2(x,y:\mathbf{\boldsymbol \theta})dxdy}{A \sigma^2}
\end{equation}
takes a predetermined value. For the amplify-and-forward channel,  the variance $\eta^2$ of the noise in the transmission channels is selected such that the total SNR in the channels defined as
%\vspace{-0.2 cm}
\begin{equation}
\mathsf{SNR}_C = \frac{\int \int_A G^2( x, y: \boldsymbol \theta) dx dy}{A \eta^2} + \frac{\sigma^2}{\eta^2} \label{eq:SNR_C_ANALOG}
\end{equation}
is also set to a predefined value. For the quantize-and-forward channel, each sensor observation is quantized to one of $M$ levels using a uniform deterministic quantizer. $K$ parallel white Gaussian noise channels add samples of noise with variance $\eta^2$  selected to set the total SNR, that is defined as
%\vspace{-0.1 cm}
\begin{equation}
\mathsf{SNR}_C = \frac{ {\int \int}_A E\left[q^2(R(x,y:\boldsymbol \theta))\right]dxdy}{A \eta^2}, \label{eq:SNR_C}
\end{equation}
to a specific value. Note that (\ref{eq:SNR_C}) converges asymptotically to (\ref{eq:SNR_C_ANALOG}) when the number of quantization levels tends to infinity.

It is assumed that the FC observes the noisy quantized samples of the field. The function $q(R(x,y:\boldsymbol \theta))$ in (\ref{eq:SNR_C}) is a quantized version of $R(x,y:\boldsymbol \theta).$
Note that due to the symmetry of the experimental set up and due to the statistical averaging, the results for $\hat{x}_c$ and $\hat{y}_c,$ and furthermore the results for $\hat{\rho}_x$ and $\hat{\rho}_y$ are very similar. Therefore, to preserve the space, convergence of solutions of iterative algorithms is demonstrated for the case of $\hat{x}_c$, $\hat{\rho}_x$ and $\hat{h}$ only.
% and for convenience in this section only the results for $\hat{x}_c$ are shown.

\begin{figure}[!t]
\centering
\includegraphics[width=9cm, height=9.5 cm]{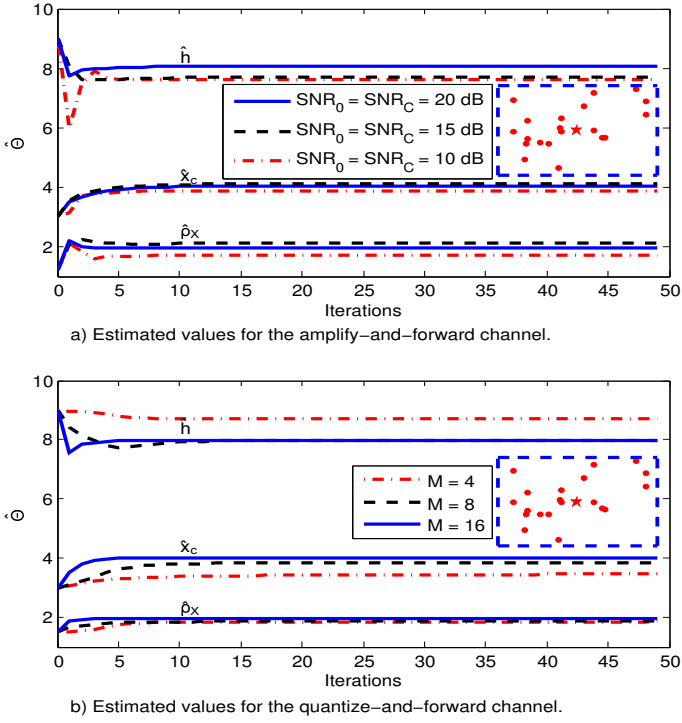}
\vspace{-0.35cm}
\caption{Estimated values of the $x$-location, the standard deviation $\rho_x$ and the strength $h$ of the field,  as a function of the number of iterations for the case of both a) the amplify-and-forward channel, and b) quantize-and-forward channel. The plots are parameterized in a) by varying values of $\mathsf{SNR}_O$ and $\mathsf{SNR}_C,$ while in b) by the number of quantization levels, $M.$ The network topologies are shown in the inset. The source of the field is represented by the star at the center of the square arena, while the K=20 sensors are shown as red dots.}\label{fig:Theta_iterations}
\vspace{-0.45cm}
\end{figure}
\vspace{-0.2 cm}
\subsection{Amplify-and-forward channel}
\label{numerical_analysis_analog_case}
First, the convergence of the ML estimator numerically evaluated by means of Newton's method is illustrated. The estimated values of the x-location, the standard deviation $\rho_x$ and the strength $h$ of the field are plotted in Fig. \ref{fig:Theta_iterations}(a) as a function of the number of iterations. The functions in Fig. \ref{fig:Theta_iterations}(a) are parameterized by different values of the SNR in the transmission and observation channels. This illustration is based on a single realization of the distributed network with $K=20$, when the initial values are picked to be $3$ for the x-location, $1.5$ for the standard deviation and $9$ for the strength of the field. Note that for the case of $\mathsf{SNR}_O=\mathsf{SNR_C}=20$ dB, the estimated values converge to the real values after 14 iterations. For the other two cases ($\mathsf{SNR}_O=\mathsf{SNR_C}=15$ dB and $\mathsf{SNR}_O=\mathsf{SNR}_C=10$ dB) the increasing discrepancy between the estimated and the true values is due to a lower sensor density in the network and also due to increasing variances of observation and transmission noise. Note how large is the deviation of the asymptotic value of $\hat{x}_c$, $\hat{\rho}_x$, and $\hat{h}$ from the real values when both $SNR_O$ and $SNR_C$ are set to $10$ dB.

To further analyze the estimation performance, the squared error (SE) between the estimated and true location parameters is evaluated. The SE is defined as
%\vspace{-0.15 cm}
\[ \mathsf{SE}  = \sum_{i=1}^{L}\left( \hat{\theta}_i-\theta_i \right)^2 \]
and the mean square error (MSE) is evaluated numerically by means of Monte Carlo simulations.

\begin{figure}[!t]
\centering
\includegraphics[width=9cm, height=9.5 cm]{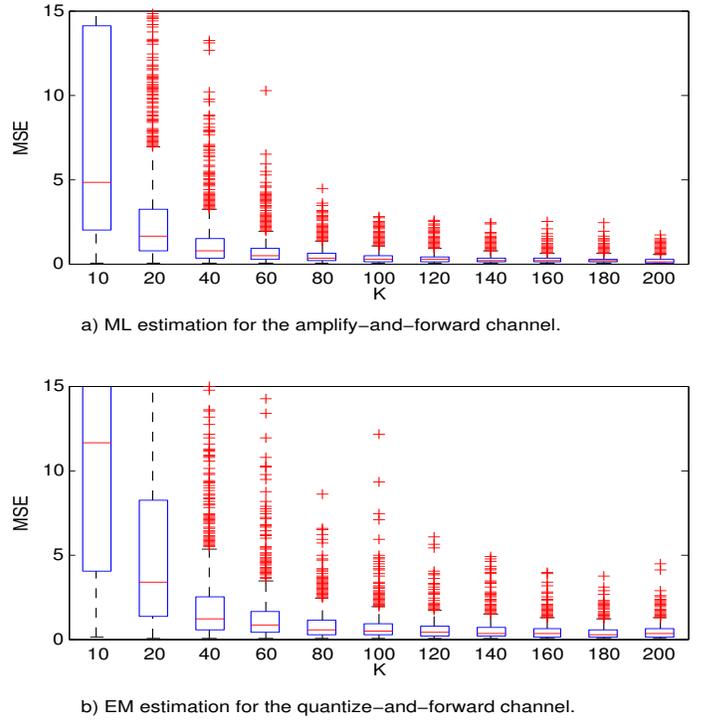}
\vspace{-0.35cm}
\caption{A box plot of the MSE between the estimated and true parameters of the field displayed as a function of the number of sensors distributed over the area $A$ for both a) the amplify-and-forward channel, and b) the quantize-and-forward channel. The SNR for the observation and transmission channel are fixed to $\mathsf{SNR}_O=\mathsf{SNR}_C=15$ dB and for the quantize-and-forward channel the number of quantization levels is set to $M=8$.}
\label{fig:MSE_K}
\vspace{0.10cm}
\end{figure}

In this and the following subsections, we involve a box plot to illustrate the dependencies of the MSE on several parameters. The central mark in each box is the median. The edges of a box present the $25$th and $75$th percentiles. The dashed vertical lines mark the data that extend beyond the two percentiles, but not considered as outliers. The outliers are then plotted individually and marked with a ``+'' sign.

\begin{figure}[!t]
\centering
\includegraphics[width=9cm, height=9.5 cm]{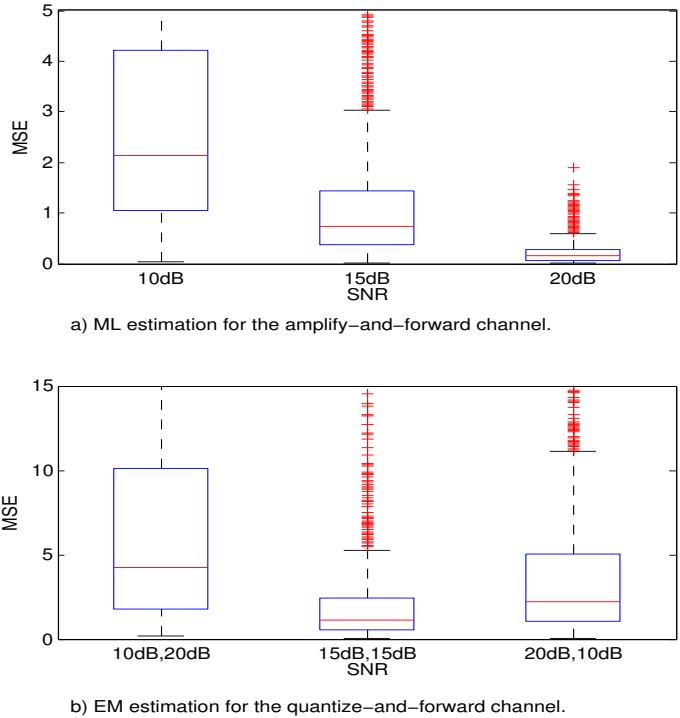}
\vspace{-0.35cm}
\caption{A box plot of the MSE between the estimated and true parameters of the field displayed as a function of the SNR for the observation and transmission channels for both a) the amplify-and-forward channel, and b) the quantize-and-forward channel. For the quantize-and-forward channel three different combinations of $\mathsf{SNR}_O$ and $\mathsf{SNR}_C$ are shown along the x-axis in the order $(\mathsf{SNR}_O,\mathsf{SNR}_C).$ The number of sensors distributed over the area $A$ is $K=40.$  The number of quantization levels is fixed to $M=8$ for the quantize-and-forward channel.}\label{fig:MSE_SNR}
\vspace{-0.15cm}
\end{figure}

\begin{figure}[!t]
\vspace{-0.05cm}
\centering
\includegraphics[width=9cm, height=9.5 cm]{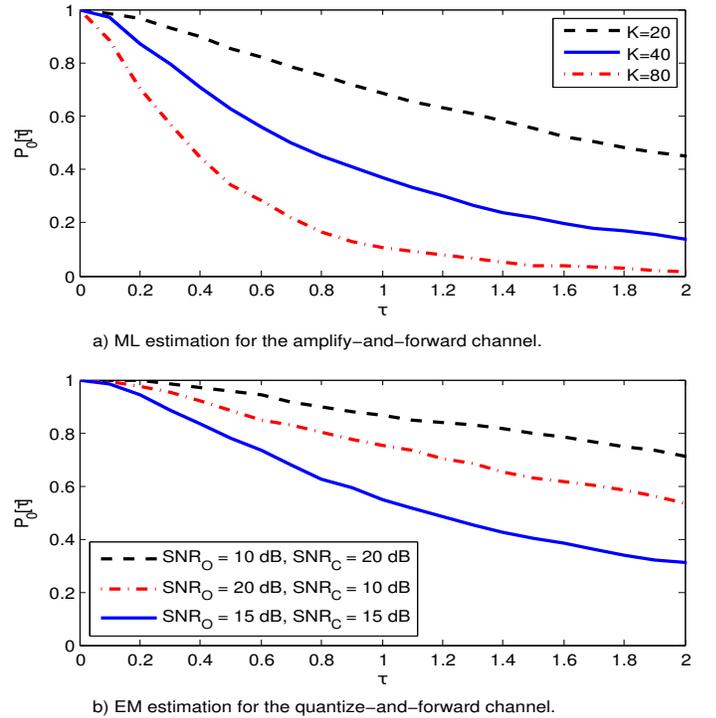}
\vspace{-0.35cm}
\caption{Probability of outliers $P_O[\tau]=P[\mathsf{SE}>\tau]$ as a function of threshold $\tau,$ $\tau>0,$ for a) the amplify-and-forward channel, and b) the quantize-and-forward channel. The curves are parameterized by different number of sensors deployed over the area $A$ for the amplify-and-forward channel. The SNR in the observation and transmission channels is fixed at $\mathsf{SNR}_O=\mathsf{SNR}_C=15$ dB. For the quantize-and-forward channel, the curves are parameterized by different values of SNR in the observation and transmission channels for a sparse network composed of $K=40$ sensors.  The number of quantization levels is set to $M=8$.}
\label{fig:P_0}
\vspace{-0.40cm}
\end{figure}

The dependence of the MSE on the number of sensors, $K,$ in the distributed network for the case of $\mathsf{SNR}_O=\mathsf{SNR}_C=15$ dB is shown as a set of box plots in Fig. \ref{fig:MSE_K}(a). The dependence of the MSE on the SNR of the observation and transmission channels, when the number of sensors is fixed to $K=40$, is displayed as a set of box plots in Fig. \ref{fig:MSE_SNR}(a). Each box in Fig. \ref{fig:MSE_K}(a) and Fig. \ref{fig:MSE_SNR}(a) is generated using $1000$ Monte Carlo realizations of the network and ML runs. To take a closer look at the distribution of outliers, we define the probability of outliers as a probability that SE exceeds a positive valued threshold $\tau.$  Denote by $P_O(\tau)$ the probability of outliers at threshold $\tau.$ Then mathematically $P_O(\tau)$ is defined as
%\vspace{-0.15 cm}
\begin{equation}
P_O(\tau) = P[\mathsf{SE}>\tau].
\label{outage_prob}
\end{equation}
We vary the value of the threshold and display the percentage of outliers as a function of $\tau$ in Fig. \ref{fig:P_0}(a). Note the large percentage of outliers for small values of $K.$ This corresponds to the case when one or more of the five parameters (x-location, y-location, the standard deviation $\rho_x$ and $\rho_y$ and the strength $h$ of the field ) did not converge to its true value.

\begin{figure}[!t]
\centering
\includegraphics[width=9cm, height=9.5 cm]{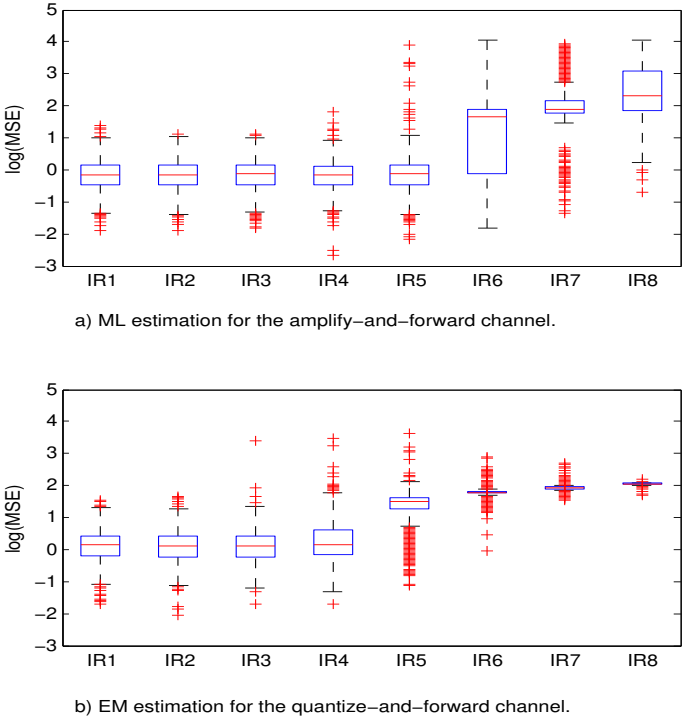}
\vspace{-0.35cm}
\caption{A box plot of the MSE between the estimated and true parameters of the field displayed as a function of the regions $\mathsf{IR}_i,$ $i=1, ... ,8,$ within which the initial values for both a) the Newton's method and (b) the EM algorithm are selected at random. The Newton's method is applied to the example with amplify-and-forward channel, while the EM algorithm is applied to the example with quantize-and-forward channel. The number of sensors is fixed to $K=40$ and the variances of noise in the observation and transmission channels are selected such that $\mathsf{SNR}_O=\mathsf{SNR}_C=15$ dB. The number of quantization levels $M$ equals to 8 for the quantize-and-forward channel.}\label{fig:IR}
\vspace{-0.25cm}
\end{figure}

The convergence of the Newton's iterations to the true values of the location parameter is analyzed with respect to a choice of initial guess for all five unknown parameters ($h, \rho_x, \rho_y, x_c, y_c$). In particular the initial values for the iterative solution due to Newton's method are selected randomly for each new Monte Carlo realization and for each unknown parameter within eight regions, that we indicate with $\mathsf{IR}_i,$ $i=1, ... ,8.$
We have enumerated these regions starting from the one closest to the true values. In particular the possible initial values for the $j-th$ unknown parameter, that can be chosen inside the $i$-th region, are in the interval $[t_j \left( 1-\frac{i-1}{8} \right), t_j \left( 1-\frac{i}{8}\right) ]$, where $t_j$ is the true value for the $j-th$ parameter.
Fig. \ref{fig:IR}(a) shows a box plot of the MSE between the estimated and true parameters of the field displayed as a function of the regions $\mathsf{IR}_i,$ $i=1, ... ,8,$ within which the initial values for the Newton's method are selected at random.
Each box plot is obtained by using $1000$ Monte Carlo realizations of the network and ML runs. The number of sensors is fixed and equal to $K=40.$ The variance $\sigma^2$ and $\eta^2$ are chosen such that the SNR for the observation and transmission channels is each equal to 15 dB. Fig. \ref{fig:IR}(a) can be interpreted as a sensitivity analysis of the implemented ML solution due to Newton's method.  As Fig. \ref{fig:IR}(a) demonstrates, it is still possible to estimate the vector of unknown parameters $\boldsymbol \theta$ with a relatively low value of MSE, even when the initial values are selected far apart from the true values, but after a certain region the Newton's method is not able to converge anymore and the MSE starts to increase rapidly. This effect is due to the presence of multiple local maxima in the log-likelihood function given by (\ref{eq:complete_log}).

\begin{figure}[!t]
\centering
\includegraphics[width=9cm, height=9.5 cm]{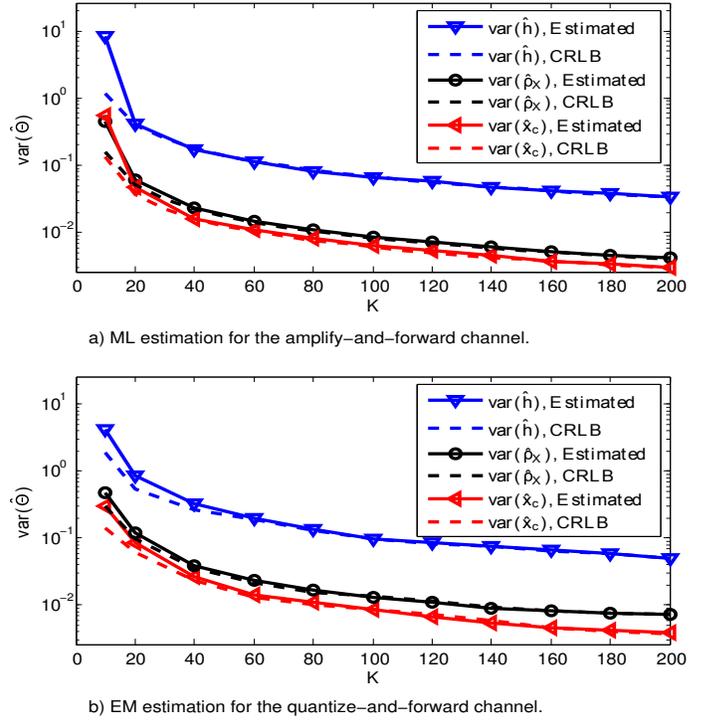}
\vspace{-0.35cm}
\caption{Variance of the estimated parameters $\hat{x}_c$, $\hat{h}$ and $\hat{\rho}_x$ and the CRLB for both a) the amplify-and-forward channel and b) the quantize-and-forward channel as a function of the number of sensors when $\mathsf{SNR}_O=\mathsf{SNR}_C= 15$ dB. The number of quantization levels is fixed to $M=8$ for the quantize-and-forward channel.}\label{fig:CRLB}
\vspace{0.9cm}
\end{figure}

Finally, Fig. \ref{fig:CRLB}(a) shows the variance of the estimated parameters $\hat{x}_c$, $\hat{h}$, and $\hat{\rho}_x$ obtained numerically by means of the Newton's method.
%by solving the maximum likelihood estimation using the Newton's method
The results are averaged over $1000$ Monte Carlo realizations. The plots are compared to the CRLB displayed as a function of the number of sensors when $\mathsf{SNR}_O=\mathsf{SNR}_C=15$ dB. The results in Fig. \ref{fig:CRLB}(a) demonstrate that for the amplify-and-forward channel, the empirical variance of the estimated parameters converges to the values obtained by means of the CRLB, underlining that with only a few distributed measurements ($K=40$) the Newton's iterative solution is efficient.

\vspace{-0.25cm}
\subsection{Quantize-and-forward channel}
\label{sec:Digital channel}
Fig. \ref{fig:Theta_iterations}(b) shows the estimated values obtained by the ML estimator in (\ref{eq:structure}) for the $x$-location, the standard deviation $\rho_x$
and the strength $h$ of the field as a function of the number of EM iterations. Different numbers of quantization levels have been considered. For this plot, a single realization of a sparse distributed network composed of $K=20$ is considered, the initial values are picked to be $3$ for the $x$-location, $1.5$ for the standard deviation and $9$ for the strength
of the field. The SNR for the observation and transmission channel is fixed to $\mathsf{SNR}_O=\mathsf{SNR}_C=15$ dB. Fig. \ref{fig:Theta_iterations}(b) illustrates the convergence of the EM algorithm, pointing out the role of the quantization levels: when $M=16$ the EM algorithm converges to the true values after only few iterations. For the other two cases, the EM algorithm does not converge as fast, and there is a larger discrepancy between the estimated and the true values of the parameters $x_c$, $\rho_x$ and $h$. This discrepancy grows as the number of quantization levels $M$ decreases, as it is expected.
%it can be noticed a discrepancy between the vectors of estimated and true parameters, that increases decreasing the number of quantization levels.
In addition to rough quantization, it is affected by the low sensor density in the network and by the noise in observation and transmission channels.

In order to analyze the performance of the estimator, in the same way it has been done for the amplify-and-forward channel, the MSE between the estimated and true parameters is used as a performance metric. Fig. \ref{fig:MSE_K}(b) shows the dependence of the MSE on the number of sensors, $K,$ in the distributed network for the case of $\mathsf{SNR}_O=\mathsf{SNR}_C=15$ dB. Fig. \ref{fig:MSE_SNR}(b) demonstrates the dependence of MSE on the varying values of SNR in the observation and transmission channels. The results are shown for the case of $M=8.$ Fig. \ref{fig:MSE_SNR}(b) is generated considering a sparse network composed of $K=40$ sensors. All three plots are generated using $1000$ Monte Carlo realizations of the network and EM runs. Fig. \ref{fig:MSE_SNR}(b) indicates that the noise in observation channel (expressed as $\mathsf{SNR}_O$) prevails over the noise in transmission channel (expressed as $\mathsf{SNR}_C$) in terms of its effect on the estimation error (the average SE and its variance). %points out that the effect of the SNR in the transmission channel is more pronounced compared to the effect of the SNR in the observation channel. In particular, the SE increases much more when the transmission channel becomes noisier than the observation channel.

The percentage of outliers due to divergence of the EM algorithm is illustrated in Fig. \ref{fig:P_0}(b). A sparse network composed of $K=40$ sensors is considered for $M=8$ and three different combinations of $SNR_O$ and $SNR_C$ are analyzed. The three combinations are (1) $\mathsf{SNR}_O=10$ dB $ and $ $\mathsf{SNR}_C=20$ dB, (2) $\mathsf{SNR}_O=\mathsf{SNR}_C=15$ dB, and (3) $\mathsf{SNR}_O=20$ dB $ and $ $\mathsf{SNR}_C=10$ dB. Fig. \ref{fig:P_0}(b) highlights that the observation channel not only affects more the performance in terms of SE than the transmission channel, but it also affects the convergence of the EM algorithm in a more tangible way, increasing the probability of outliers. This is a consequence of quantizing very noisy measurements.
%This effect can be attributed to the noisier measurements that are furthermore deteriorated by the rough quantization.

\begin{figure}[t]
\centering
\includegraphics[width=9cm]{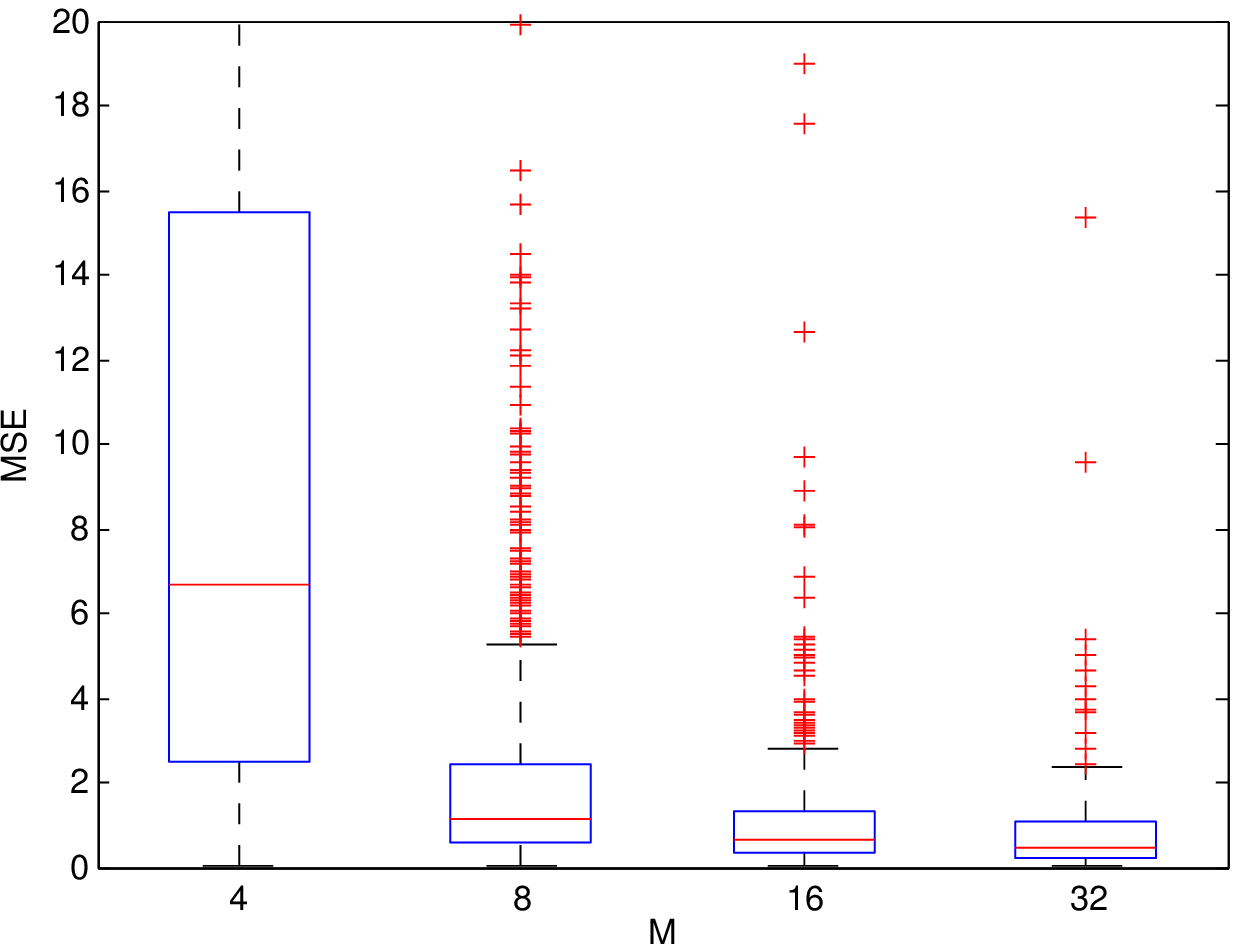}
\vspace{-0.75cm}
\caption{A box plot of the MSE between the estimated and true parameters of the field displayed as a function of the number of quantization levels for a sparse network composed of $K=40$ sensors. The SNR in the observation and transmission channels is fixed to $15$ dB.  }
\label{fig:DigCh_SE_M}
\vspace{0.55cm}
\centering
\includegraphics[width=9cm]{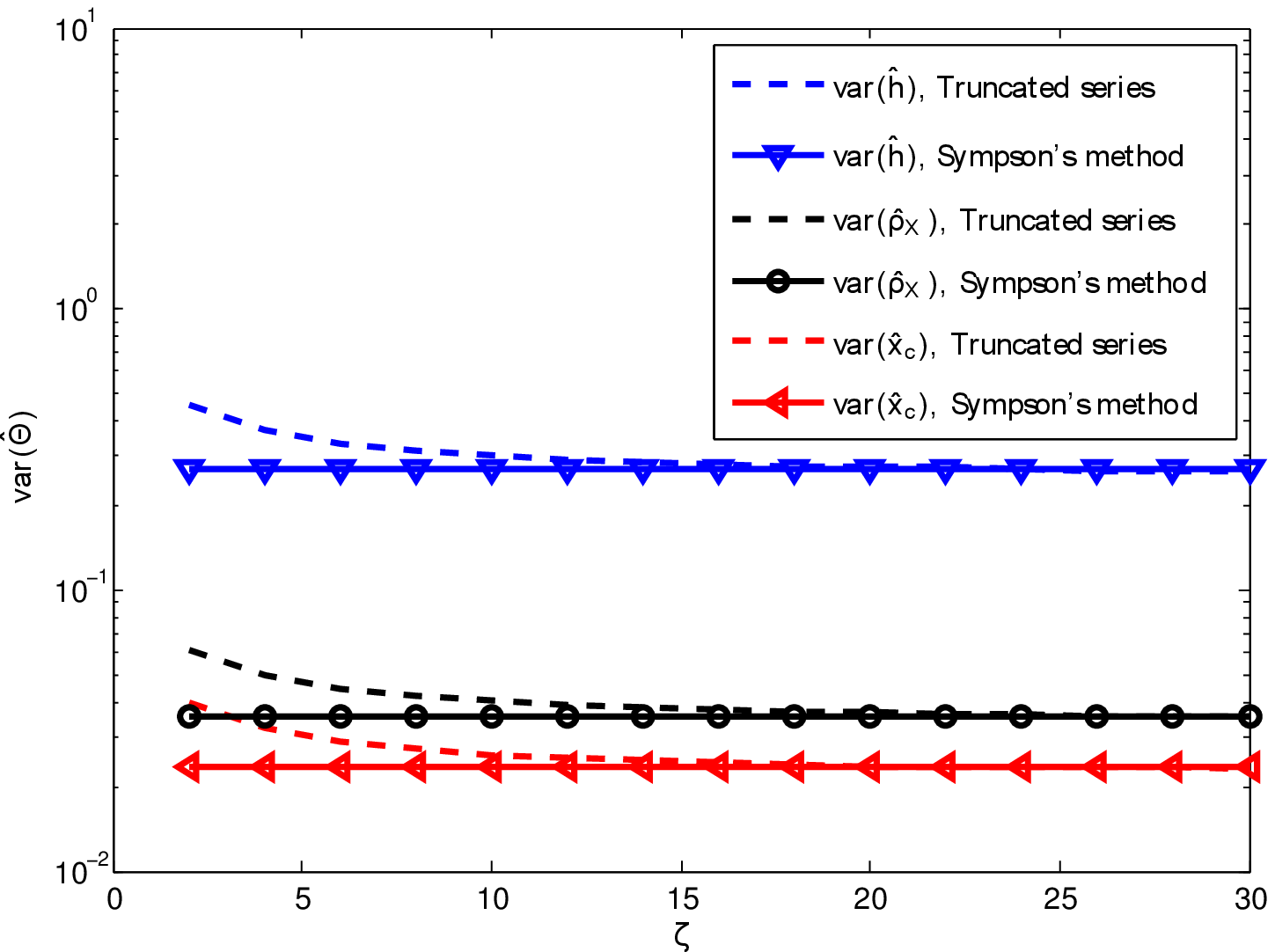}
\vspace{-0.60cm}
\caption{CRLB of the estimated parameter $\hat{x}_c$, $\hat{\rho}_x$ and $\hat{h}$ as function of the number of terms $\zeta$ used to truncate the infinite series in (\ref{FinalExpression}) compared with the case when Simpson's method is used. The number of quantization levels is fixed to $M=8$, the number of sensors are fixed to $K=40$ and the SNRs are set to $\mathsf{SNR}_O=\mathsf{SNR}_C=15$ dB.}\label{fig:truncation}
\vspace{-0.55cm}
\end{figure}

In order to analyze the robustness of the EM algorithm to the initial values of estimates, eight regions $\{ \mathsf{IR}_1, \mathsf{IR}_2,...,\mathsf{IR}_8\}$, inside which the initial values are chosen randomly, are considered. In particular the possible initial values for the $j-th$ unknown parameter, that can be chosen inside the $i$-th region, are in the interval $[t_j \left( 1-\frac{i-1}{8} \right), t_j \left( 1-\frac{i}{8}\right) ]$, where $t_j$ is the true value for the $j-th$ parameter.
Fig. \ref{fig:IR}(b) shows a box plot of the MSE between the estimated and true parameters of the field displayed as a function of the regions $\mathsf{IR}_i,$ $i=1, ... ,8,$ within which initial values of parameters are selected randomly. Each box is generated by using $1000$ Monte Carlo realizations of the network and EM runs. Here the network is composed of $K=40$ sensors, the number of quantization levels is set to $M=8$, and the variances $\sigma^2$ and $\eta^2$ are selected such that the SNR for the observation and transmission channels is equal to 15 dB. Fig. \ref{fig:IR}(b) demonstrates that the EM algorithm is not very sensitive to the choice of the initial value of estimates.
Note that up to the $4$-th region the median value of MSE is reasonably low. The abrupt increment of the MSE is caused by the presence of multiple local maxima in the log-likelihood function given by (\ref{eq:incomplete_likelihood}).

Fig. \ref{fig:DigCh_SE_M} shows the effect of the number of quantization levels on the MSE,  when the variance $\sigma^2$ and $\eta^2$ are chosen such that the SNR for both observation and transmission channels is set to $15$ dB. As expected the MSE decreases when the number of quantization levels is increased. As the number of quantization levels at local sensors increases, the percentage of this performance improvement decreases and tends to converge to the case when raw observations are transmitted. Nevertheless the distributed parameter estimation system achieves an acceptable performance in terms of MSE even when $M=8$ and for a sparse network composed of $K=40$ sensors. This emphasizes on the energy efficiency of the proposed parameter estimation framework that could lead to a higher lifetime of distributed sensors in the network. In other words, local sensors do not need to waste a lot of energy to send high-resolution quantized observations to achieve an acceptable estimation performance in terms of the MSE.

Fig. \ref{fig:CRLB}(b) shows the variance of the estimated parameter $\hat{x}_c$, $\hat{\rho}_x$ and $\hat{h}$, obtained by means of the EM algorithm for the case of $M=8.$  The results are averaged over 1000 Monte Carlo realizations.  The plots are compared to the CRLB displayed as a function of the number of sensors. Fig. \ref{fig:CRLB}(b) demonstrates that for the quantize-and-forward channel, the EM solution is efficient.  The convergence rate of variance of the estimated parameters to the value provided by the CRLB is slightly slower compared to the similar plots in the case of the amplify-and-forward channel.

Lastly, Fig. \ref{fig:truncation} shows the dependence of the truncated CRLB  on the number of terms $\zeta$ retained in the infinite series  (\ref{FinalExpression}). The results of truncation are compared to those obtained by means of numerical integration using the Simpson's method.  The plots are obtained for the case of $M=8$, $K=40$ and $\mathsf{SNR}_O=\mathsf{SNR}_C=15$ dB. Fig. \ref{fig:truncation} shows that by truncating the infinite series in (\ref{FinalExpression}), it is possible to achieve the same values of the CRLB as those achieved by the Simpson's method.

%required accuracy is achieved and (\ref{Fisher_eq6}) is evaluated with an acceptable error, that can be computed
%by
%\begin{eqnarray}
%\epsilon =  \sum_{n=\tau}^{\infty } \hspace{-0.05cm} \sum_{m=0}^n \hspace{-0.05cm} \left( -1\right)^{m+n}\hspace{-0.05cm}  \frac{n!}{m!} \hspace{-0.65cm} \mathop{ \sum_{\ell_v \geq 0}}_{\sum_{v=0}^{M}\ell_v=n-m}
% \hspace{-0.75cm} \left(\prod_{v=1}^M \hspace{-0.05cm} \frac{ p_{k,v}\hspace{-0.05cm} (\boldsymbol \theta) }{\ell_v !} \hspace{-0.05cm} \right)\hspace{-0.05cm} \Lambda_{\ell_v}.
%\label{error}
%\end{eqnarray}
\vspace{-0.25cm}
\subsection{Comparison Between EM and NR}
\label{sec:Comparison Between EM and NN}

\begin{figure}[t]
\centering
\includegraphics[width=9cm]{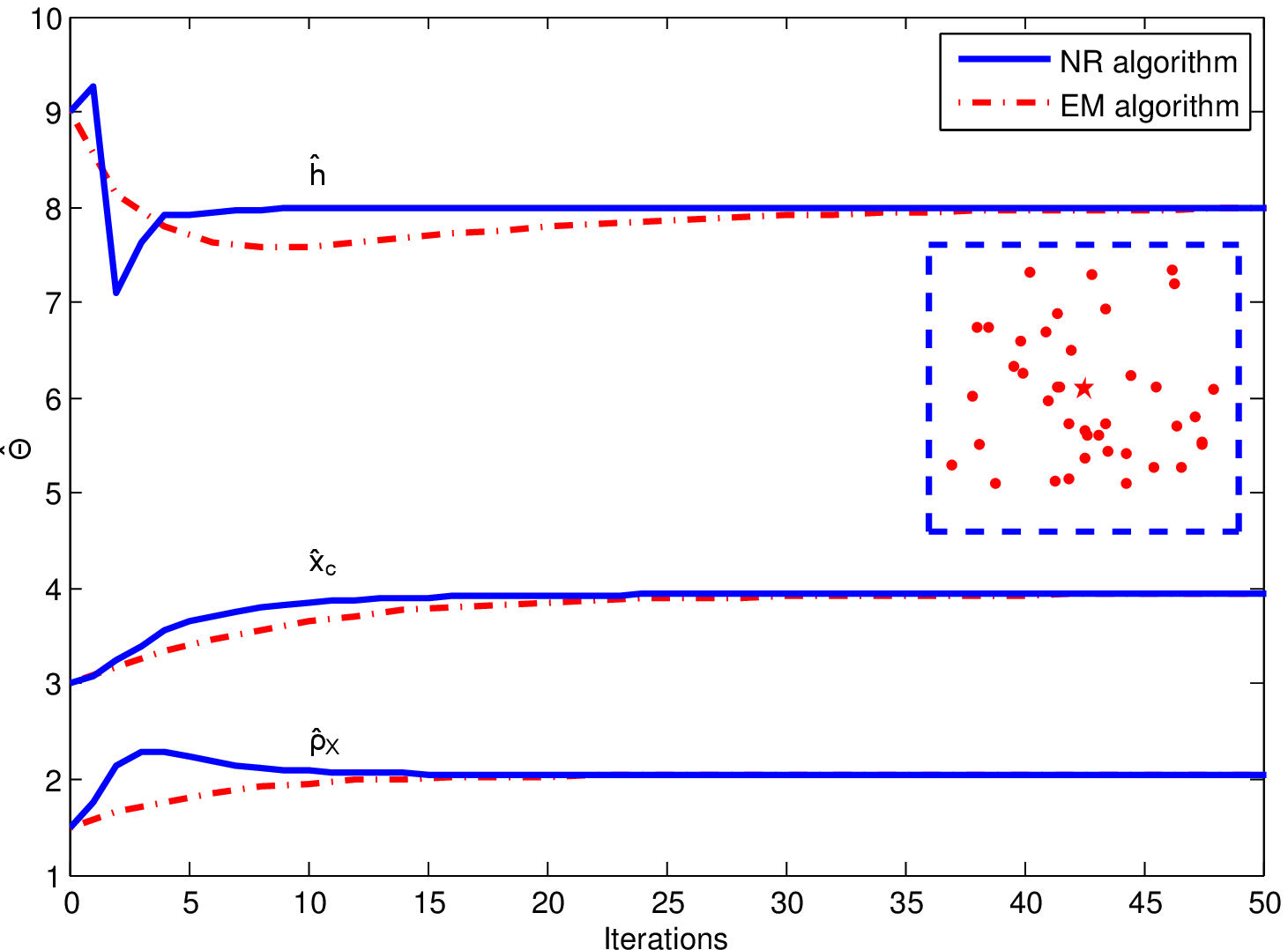}
\vspace{-0.55cm}
\caption{Illustration of EM and NR-iterations for estimation of x-location, the standard deviation $\rho_x$ and the strength $h$ of the field. The SNR ratio in the observation and transmission channels is set to $15$ dB. The network topology
is shown in the inset. The source of the field is represented by the star at the center of the square arena, while the K=40 sensors are shown as red dots.}\label{fig:Theta_EMvsNR}
\vspace{0.25cm}
\includegraphics[width=9cm]{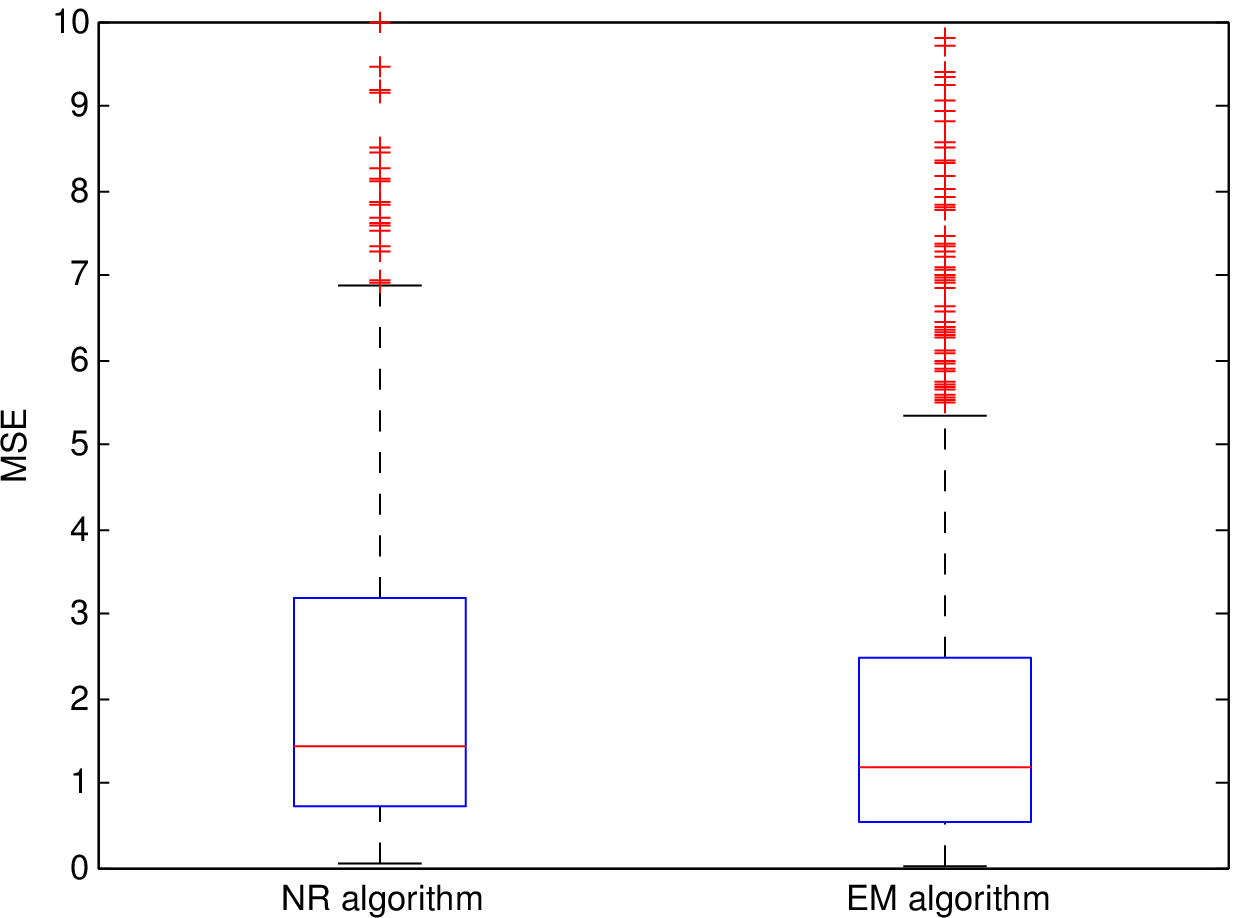}
\vspace{-0.60cm}
\caption{A box plot of the MSE between the estimated and true parameters of the field displayed for both the NR and the EM algorithm. The number of sensors is fixed to $K=40.$ The SNRs are set to $\mathsf{SNR}_O=\mathsf{SNR}_C=15$ dB. The number of quantization levels $M$ is set to 8.}\label{fig:MSE_EMvsNR}
\vspace{-0.05cm}
\end{figure}

Fig. \ref{fig:Theta_EMvsNR} shows the estimated values for the $x$-location, the standard deviation $\rho_x$ and the strength $h$ of the field as a function of the number of iterations for both the EM and the NR algorithms. For this plot, a single realization of a sparse sensor network composed of $K=40$ is considered, the initial values are picked to be $3$ for the $x$-location, $1.5$ for the standard deviation and $9$ for the strength
of the field for both algorithms. The SNR in the observation and transmission channels is fixed to $\mathsf{SNR}_O=\mathsf{SNR}_C=15$ dB, and the number of quantization levels is fixed to $M=8$. Fig. \ref{fig:Theta_EMvsNR} shows that when both algorithms converge, the NR algorithm, as stated in Sec. \ref{sec:Alternative solution}, requires fewer iterations to reach the final values compared to the EM algorithm, underlining the advantage of the NR algorithm in terms of convergence rate.

\begin{figure}[t]
\centering
\includegraphics[width=9cm]{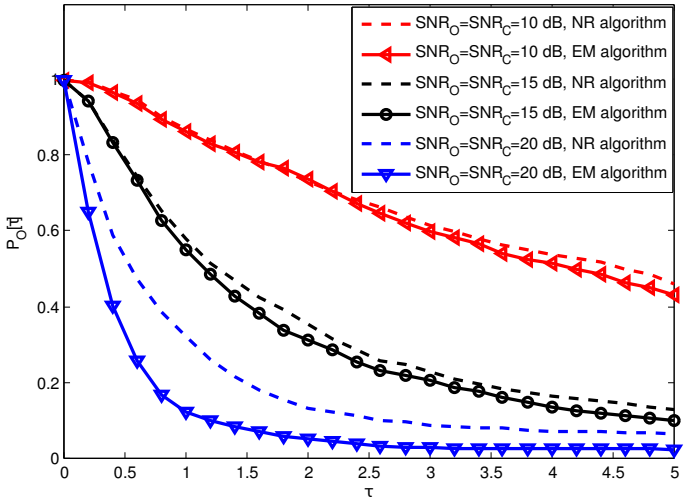}
\vspace{-0.75cm}
\caption{Probability of outliers $P_O[\tau]=P[\mathsf{SE}>\tau]$ as a function of $\tau$ for both EM and NR algorithms parameterized by different values of SNR in the observation and transmission channels. The network is composed of $K=40$ sensors and the number of quantization levels is set to $M=8$. }\label{fig:P0_EMvsNR}
\vspace{-0.25cm}
\end{figure}

In order to compare the two algorithms in terms of their accuracy, the MSE between the estimated and the true parameters is evaluated.  Fig. \ref{fig:MSE_EMvsNR} shows the MSE for both EM and NR algorithms.  For this example, the network is composed of $K=40$ sensors, the SNRs are set to  $\mathsf{SNR}_O=\mathsf{SNR}_C=15$ dB and $M=8$.  Fig. \ref{fig:MSE_EMvsNR} shows that the EM algorithm has lower MSE compared to the NR.  To further analyze the performance of the algorithms, Fig. \ref{fig:P0_EMvsNR} displays the percentage of outliers as a function of the threshold $\tau$ (see equation (\ref{outage_prob})) for both algorithms and for three different combinations of $\mathsf{SNR}_O$ and $\mathsf{SNR}_C$: (1) $\mathsf{SNR}_O=\mathsf{SNR}_C=10$ dB, (2) $\mathsf{SNR}_O=\mathsf{SNR}_C=15$ dB, and (3) $\mathsf{SNR}_O=\mathsf{SNR}_C=20$ dB.

\vspace{-0.35cm}
\section{Summary}
\label{sec:Summary}
This paper has presented a distributed ML estimation procedure based on an iterative solution for estimating a parametric physical field. Furthermore, it has also detailed the derivation of a transformed expression for the CRLB on the variance of distributed estimates of field parameters by a homogeneous sensor network. The model of the network assumed independent sensor measurements and transmission over a noisy environment. Two channel models were considered: (1) the measurements from the sensors were sent directly to the FC by means of a linear analog modulation; (2) each sensor quantized its measurement to $M$ levels and the quantized and encoded data were communicated to the FC over parallel additive white Gaussian channels.
The stability of the distributed parameter estimator has been analyzed for both models and also its robustness to the initial values of estimates has been considered. The results have shown that for the quantize-and-forward channel the SNR of the observation channel dominates the SNR of the transmission  channel in terms of the values of MSE and it also affects the convergence of the EM algorithm in a more tangible way, increasing the probability of outliers.
%In order to do a complete analysis of the performance of the estimator, the effect of the number of quantization levels on the SE has been analyzed.
The developed CRLBs were further compared to the numerical values of the variance of estimates.
%used to evaluate the performance of the estimates of the field parameters for both channels.
The results showed that the estimates are nearly efficient for small $K$ and become efficient for both channel models when the density of the sensor network increases.  We have demonstrated numerically that the derived EM algorithm generates more accurate estimates compared to the NR method.  Finally, we have shown that truncating the infinite sum in the developed CRLB to only few first terms produces highly accurate results.
%The variance of the estimated parameters approaches the CRLB.

As a future work, the parametric field will be replaced by a mixture model or nonparametric field (for generality), mimicking the case of unknown number of multiple objects generating a cumulative physical field sensed by a distributed sensor network within an area $A.$  The models for transmission channels will involve fading and shadowing effects. Efforts to eliminate the FC and make the sensor network decentralized will be made.

\vspace{-0.25cm}
\appendices
\section{}
This section provides details leading to the equation (\ref{eq:structure}). Consider the $k$-th term under the sum in (\ref{eq:M-step}):
\[ E\left[ \left. (R_k-G_k)\frac{\partial G_k}{\partial \theta_t} \right| \mathbf{z}_k,\hat{\mathbf{\boldsymbol \theta}}^{(m)} \right]   \]
%\vspace{-0.25cm}
\[ = \int_{-\infty}^{+\infty} (r_k-G_k) \frac{\partial G_k}{\partial \theta_t} \frac{\exp\left( -\frac{(r_k-G_k^{(m)})^2}{2\sigma^2_k}\right)}{f_{\mathbf{Z}_k}^{(m)}(\mathbf{z}_k) \sqrt{2\pi\sigma^2_k}} \]
%\vspace{-0.15cm}
\[ \times \frac{\exp\left(-\frac{(\mathbf{z}_k-\mathbf{B}^{(m)}_k)^T(\mathbf{z}_k-\mathbf{B}^{(m)}_k)}{2 \eta^2_k}\right)}{\left( 2 \pi \eta^2_k\right)^{\alpha/2}} dr_k \]
%\vspace{-0.25cm}
\[ = \sum_{j=1}^M \int_{\tau_j}^{\tau_{j+1}} (r_k-G_k) \frac{\partial G_k}{\partial \theta_t} \frac{\exp\left(-\frac{(r_k-G_k^{(m)})^2}{2\sigma^2_k}\right)}{f_{\mathbf{Z}_k}^{(m)}(\mathbf{z}_k)  \sqrt{2\pi\sigma^2_k}} \]
%\vspace{-0.15cm}
\[ \times \frac{\exp\left(-\frac{(\mathbf{z}_k-\mathbf{b}_j)^T(\mathbf{z}_k-\mathbf{b}_j)}{2 \eta^2_k}\right)}{\left( 2 \pi \eta^2_k\right)^{\alpha/2}} dr_k \]
%\vspace{-0.15cm}
\[ = \sum_{j=1}^M \frac{\exp\left(-\frac{(\mathbf{z}_k-\mathbf{b}_j)^T(\mathbf{z}_k-\mathbf{b}_j)}{2 \eta^2_k}\right)}{f_{\mathbf{Z}_k}^{(m)}(\mathbf{z}_k) \left( 2 \pi \eta^2_k\right)^{\alpha/2}} \frac{\partial G_k}{\partial \theta_t} \]
%\vspace{-0.25cm}
\[ \times \int_{\tau_j}^{\tau_{j+1}} (r_k-G_k)\frac{\exp\left(-\frac{(r_k-G_k^{(m)})^2}{2\sigma^2_k}\right)}{\sqrt{2\pi\sigma^2_k}} dr_k.\]

Note that the difference $(r_k-G_k)$ in the last integral can be rewritten as $(r_k-G_k^{(m)}+G_k^{(m)}-G_k).$ Then
%\vspace{-0.15cm}
\small
\[E\left[ \left. (R_k-G_k)\frac{\partial G_k}{\partial \theta_t} \right| \mathbf{z}_k,\hat{\mathbf{\boldsymbol \theta}}^{(m)} \right] \hspace{-0.1 cm}= \hspace{-0.1 cm} \sum_{j=1}^M \frac{\exp\left(-\frac{(\mathbf{z}_k-\mathbf{b}_j)^T(\mathbf{z}_k-\mathbf{b}_j)}{2\eta^2_k}\right)}{f_{\mathbf{Z}_k}^{(m)}(\mathbf{z}_k) \left( 2 \pi \eta^2_k\right)^{\alpha/2}} \frac{\partial G_k}{\partial \theta_t} \]
\[ \times \left\{ \frac{1}{\sqrt{2 \pi \sigma^2_k}} \int_{\tau_j}^{\tau_{j+1}} \exp \left( -\frac{(r_k-G_k^{(m)})^2}{2\sigma^2_k} \right) d \frac{(r_k-G^{(m)}_k)^2}{2} \right. \]
%\vspace{-0.15cm}
\[ \left. + (G^{(m)}_k-G_k) \frac{1}{\sqrt{2\pi \sigma^2_k}} \int_{\tau_j}^{\tau_{j+1}} \exp \left(-\frac{(r_k-G_k^{(m)})^2}{2 \sigma^2_k}\right) dr_k \right\}. \]
\normalsize
Replacing the last integral with a difference of two Q-functions we obtain:
%\vspace{-0.15cm}
\small
\[ \sum_{k=1}^K E \hspace{-0.1 cm}\left[ \hspace{-0.1 cm}\left. (R_k-G_k)\frac{\partial G_k}{\partial \theta_t} \right| \mathbf{z}_k,\hat{\mathbf{\boldsymbol \theta}}^{(m)} \right] \hspace{-0.1 cm}= \hspace{-0.1 cm}\sum_{k=1}^K \hspace{-0.1 cm} \sum_{j=1}^{M} \frac{\exp\left(-\frac{(\mathbf{z}_k-\mathbf{b}_j)^T(\mathbf{z}_k-\mathbf{b}_j)}{2\eta^2_k}\right)}{f_{\mathbf{Z}_k}^{(m)}(\mathbf{z}_k) \left( 2 \pi \eta^2_k\right)^{\alpha/2}} \]
%\vspace{-0.05cm}
\[\times \frac{\partial G_k}{\partial \theta_t} \left\{ \frac{\sigma^2_k}{\sqrt{2 \pi \sigma^2_k}} \left\{ \exp \left( -\frac{(\tau_j-G_k^{(m)})^2}{2\sigma^2_k}\right) \right. \right. \]
%\vspace{-0.05cm}
\[ \left. \left. - \exp \left(-\frac{(\tau_{j+1}-G_k^{(m)})^2}{2\sigma^2_k}\right) \right\} + (G_k^{(m)}-G_k)  \right. \]
%\vspace{-0.05cm}
\[ \left.  \left. \left\{ Q\left(\frac{\tau_j-G_k^{(m)}}{\sigma_k}\right)- Q\left(\frac{\tau_{j+1}-G_k^{(m)}}{\sigma_k}\right) \right\} \right\} \right|_{G_k=G_k^{(m+1)}} = 0.\]
\normalsize
\section{}
This section provides details leading to a closed form expression for the expectation in (\ref{Fisher_eq6}).

After some basic manipulations the right side of (\ref{Fisher_eq6}) can be represented in the following form:
%\vspace{-0.10cm}
\small
\begin{eqnarray}
\hspace{-0.4cm}\Phi_{k,j,i}\hspace{-0.05cm} &=&\hspace{-0.15cm} \underset{\mathbb R^{\alpha}}{ \int ... \int}   \frac{1}{\left( 2 \pi \eta^2_k\right)^{\alpha/2}} \nonumber \\  \hspace{-0.05cm}& &\hspace{-0.15cm} \times \frac{\exp \left( -\frac{(\mathbf{z}_k-\mathbf{b}_j)^T(\mathbf{z}_k-\mathbf{b}_j) + (\mathbf{z}_k-\mathbf{b}_i)^T(\mathbf{z}_k-\mathbf{b}_i)}{2 \eta^2_k} \right)}{\hspace{-0.05cm}\sum_{v\hspace{-0.05cm}=\hspace{-0.05cm}1}^{M} p_{k,v}\hspace{-0.05cm}\left( \boldsymbol \theta
\right) \exp \hspace{-0.1 cm}\left( \hspace{-0.05cm} -\frac{(\mathbf{z}_k-\mathbf{b}_v)^T(\mathbf{z}_k-\mathbf{b}_v)}{2 \eta^2_k} \right)} d\mathbf{z}_k. \label{Fisher_eq8}
\end{eqnarray}
\normalsize
The denominator in (\ref{Fisher_eq8}) can be further expressed as a series (see page 15 in \cite{Abramowitz:1965}):
%\vspace{-0.10cm}
\begin{eqnarray}
\left((x(\boldsymbol \theta)-1)+1 \right)^{-1} &=& \sum_{n=0}^{\infty } \left( -1\right)^n (x(\boldsymbol \theta)-1)^n.
  \label{Identity1}
\end{eqnarray}
After replacing the denominator of the integrand in (\ref{Fisher_eq8}) with the expression in the right hand side of (\ref{Identity1}), we have
%\vspace{-0.10cm}
\small
\begin{eqnarray}
\hspace{-0.50cm} \Phi_{k,j,i}\hspace{-0.3cm} & = &\hspace{-0.3cm} \sum_{n=0}^{\infty } \left( -1\right)^n \underset{\mathbb R^{\alpha}}{ \int ... \int}    \frac{1}{\left( 2 \pi \eta^2_k\right)^{\alpha/2}}  \nonumber \\
\hspace{-0.85 cm} & & \hspace{-0.75 cm}  \times \exp \left( -\frac{(\mathbf{z}_k-\mathbf{b}_j)^T(\mathbf{z}_k-\mathbf{b}_j) + (\mathbf{z}_k-\mathbf{b}_i)^T(\mathbf{z}_k-\mathbf{b}_i)}{2 \eta^2_k} \right)\nonumber \\
\hspace{-0.85 cm} & & \hspace{-0.75 cm}  \times \left[ \sum_{v=1}^{M} p_{k,v}\left(\boldsymbol \theta \right) \exp \left( -\frac{(\mathbf{z}_k-\mathbf{b}_v)^T(\mathbf{z}_k-\mathbf{b}_v)}{2 \eta^2_k} \right)-1 \right]^n \hspace{-0.2cm} d\mathbf{z}_k .\label{Fisher_eq9}
\end{eqnarray}
\normalsize
Applying the binomial theorem \cite{Abramowitz:1965}
%\vspace{-0.10cm}
\begin{eqnarray}
 \left( a+b \right)^n &=& \sum_{m=0}^n{n \choose m}a^{n-m}b^{m},
  \label{Identity2}
\end{eqnarray}
to the power of $n$ term in (\ref{Fisher_eq9}), yields:
%\vspace{-0.10cm}
\begin{eqnarray}
 \hspace{-0.25 cm}\Phi_{k,j,i} \hspace{-0.25 cm} &=& \hspace{-0.25 cm}\sum_{n=0}^{\infty } \left( -1\right)^n \sum_{m=0}^n \left \{ \left( -1\right)^m {n \choose m}  \right. \nonumber \\ \hspace{-0.85 cm} & & \hspace{-0.75 cm} \times \hspace{-0.15 cm}  \left. \underset{\mathbb R^{\alpha}}{ \int ... \int }   \frac{\exp \left( -\frac{(\mathbf{z}_k-\mathbf{b}_j)^T(\mathbf{z}_k-\mathbf{b}_j) + (\mathbf{z}_k-\mathbf{b}_i)^T(\mathbf{z}_k-\mathbf{b}_i)}{2 \eta^2_k} \right)}{\left( 2 \pi \eta^2_k\right)^{\alpha/2}} \right. \nonumber \\ \hspace{-0.85 cm} & & \hspace{-0.75 cm} \times \hspace{-0.15 cm} \left. \left[ \sum_{v=1}^{M} p_{k,v}\hspace{-0.1 cm}\left( \boldsymbol \theta\right) \hspace{-0.1 cm} \exp \hspace{-0.1 cm}\left( \hspace{-0.1 cm} -\frac{(\mathbf{z}_k-\mathbf{b}_v)^T(\mathbf{z}_k-\mathbf{b}_v)}{2 \eta^2_k} \right)\hspace{-0.1 cm} \right]^{n-m}  \hspace{-0.65cm} d\mathbf{z}_k  \right\} \hspace{-0.15 cm}. \label{Fisher_eq10}
\end{eqnarray}
At last, after involving the following multinomial expansion
%\vspace{-0.15cm}
 \begin{eqnarray}
 \left( \sum_{v=1}^M x_v \right)^w & = & w! \hspace{-0.30cm} \mathop{ \sum_{\ell_v \geq 0}}_{\sum_{v=0}^{M}\ell_v=w} \hspace{-0.30cm}
\left( \prod_{v=1}^M \frac{ x_v^{\ell_v} }{\ell_v !} \right),
  \label{Identity3}
\end{eqnarray}
where the summation on the right-hand side is over all indices that sum to $w,$  we obtain:
%\vspace{-0.15cm}
\begin{eqnarray}
\Phi_{k,j,i} =  \sum_{n=0}^{\infty } \hspace{-0.05cm} \sum_{m=0}^n \hspace{-0.05cm} \left( -1\right)^{m+n}\hspace{-0.05cm}  \frac{n!}{m!} \hspace{-0.65cm} \mathop{ \sum_{\ell_v \geq 0}}_{\sum_{v=0}^{M}\ell_v=n-m}
 \hspace{-0.75cm} \left(\prod_{v=1}^M \hspace{-0.05cm} \frac{ p_{k,v}\hspace{-0.05cm} (\boldsymbol \theta) }{\ell_v !} \hspace{-0.05cm} \right)\hspace{-0.05cm} \Lambda_{\ell_v},
\label{Fisher_eq11}
\end{eqnarray}
where
%\vspace{-0.25cm}
\small
\begin{eqnarray}
\Lambda_{\ell_v} & = & \underset{\mathbb R^{\alpha}}{ \int ... \int }  \frac{\exp \left( -\frac{(\mathbf{z}_k-\mathbf{b}_j)^T(\mathbf{z}_k-\mathbf{b}_j)+(\mathbf{z}_k-\mathbf{b}_i)^T(\mathbf{z}_k-\mathbf{b}_i)}{2 \eta_k^2} \right)}{ \left( 2 \pi \eta_k^2\right)^{\alpha/2}} \nonumber \\
& & \times \exp\left( \sum_{v=1}^M \ell_v(\mathbf{z}_k-\mathbf{b}_v)^T(\mathbf{z}_k-\mathbf{b}_v)\right)    d\mathbf{z}_k
\nonumber \\
& = &  \hspace{-0.1 cm}\exp \hspace{-0.1 cm}\left( \hspace{-0.1 cm} - \frac{\left(\mathbf{b}_j \hspace{-0.1 cm}+ \hspace{-0.1 cm}\mathbf{b}_i \hspace{-0.1 cm}+ \hspace{-0.1 cm} \sum_{v=1}^M \hspace{-0.1 cm}\ell_v \mathbf{b}_v\right)^T \hspace{-0.1 cm}\left(\mathbf{b}_j\hspace{-0.1 cm}+ \hspace{-0.1 cm}\mathbf{b}_i \hspace{-0.1 cm}+ \hspace{-0.1 cm} \sum_{v=1}^M \hspace{-0.1 cm} \ell_v \mathbf{b}_v\right) }{{2 \eta^2_k
} \sum_{v=1}^M \ell_v +2}  \hspace{-0.1 cm}\right) \nonumber \\
 & & \times \frac{\exp\left( -\frac{\mathbf{b}_j^T \mathbf{b}_j + \mathbf{b}_i^T \mathbf{b}_i + \sum_{v=1}^M \left(\mathbf{b}_v^T \mathbf{b}_v \ell_v \right)}{2 \eta^2_k
}\right)}{\sqrt{\sum_{v=1}^M \ell_v+2 }}. \label{Fisher_eq12}
\end{eqnarray}
\normalsize

\bibliographystyle{ieeetr}
\bibliography{./References}

 \nocite{Mohammadi13}
% \nocite{Kar11}
% \nocite{Tichavsky98}
 \nocite{Gupta2011}
% \nocite{Deuflhard2004}
% \nocite{Dargie2010}
% \nocite{Malson06}
% \nocite{Stoica90}
\vspace{-1.3 cm}
  \begin{IEEEbiography}[{\includegraphics[width=1in,height=1.35in,clip,keepaspectratio]{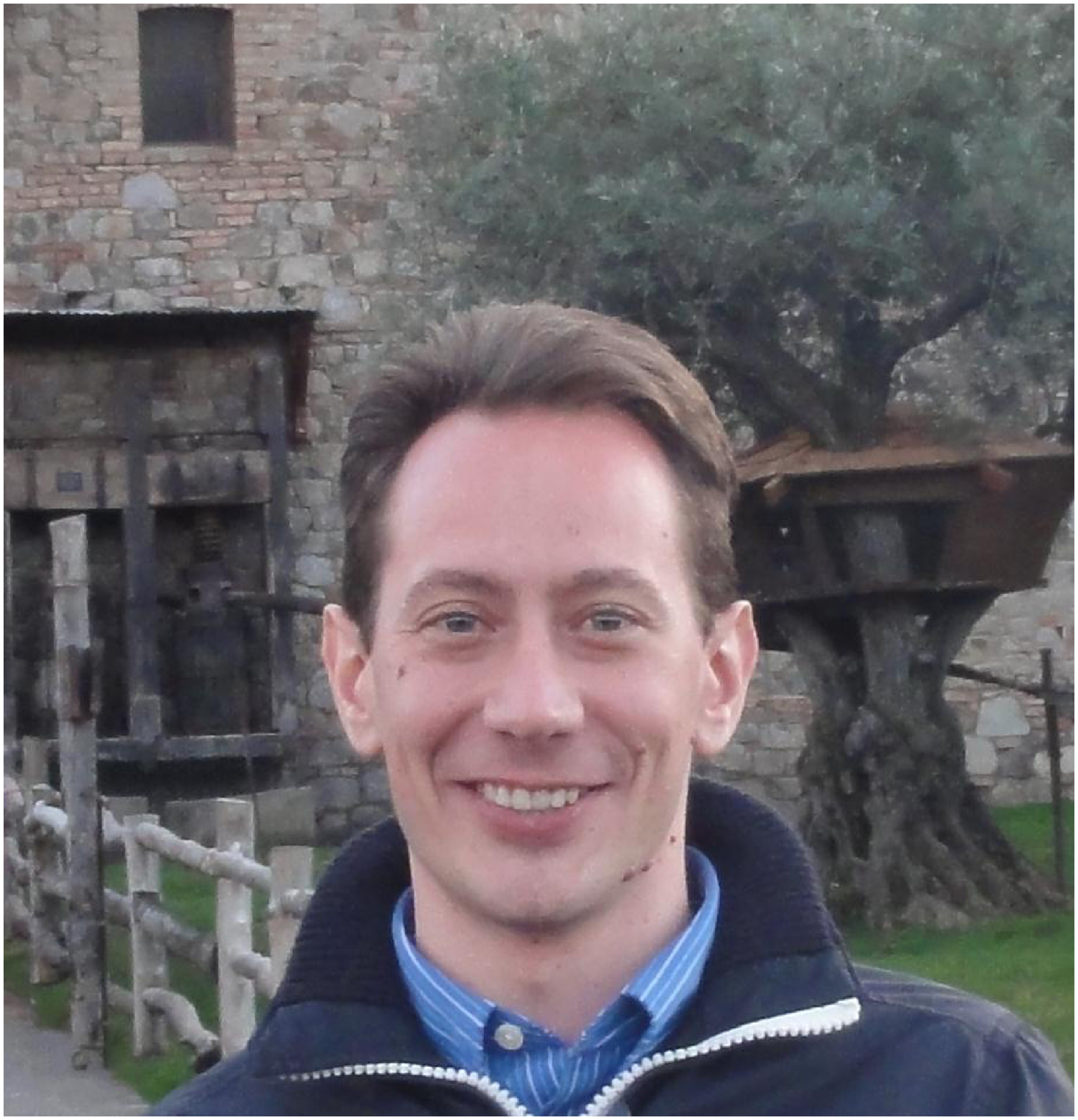}}]{Salvatore Talarico}
received the BSc and MEng degrees in electrical engineering from University of Pisa, Italy, in 2006 and 2007 respectively. From 2008 until 2010, he worked in the R$\&$D department of Screen Service Broadcasting Technologies (SSBT) as an RF System Engineer. He is currently a research assistant and a Ph.D. student in the Lane Department of Computer Science and Electrical Engineering at West Virginia University, Morgantown, WV. His research interests are in wireless communications, software defined radio and modeling, performance evaluation and optimization of ad-hoc and cellular networks.
% using tools from stochastic geometry.
\end{IEEEbiography}
\vspace{-1 cm}
  \begin{IEEEbiography}[{\includegraphics[width=1in,height=1.25in,clip,keepaspectratio]{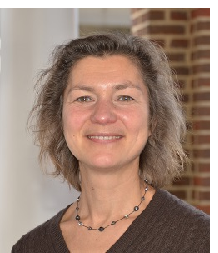}}]{Natalia A. Schmid}(S'97-M'00) received the Ph.D. degree in engineering from Russian Academy of Sciences, Moscow, Russia, in 1995 and the D.Sc. degree in electrical engineering from Washington University, St. Louis, MO, in 2000. In 2001, she was a postdoctoral research associate with the University of Illinois, Urbana. Since 2003, she has been with the Lane Department of Computer Science and Electrical Engineering, West Virginia University, Morgantown, where she is currently an associate professor. Her current research interests include detection and estimation, learning theory and information theory with application to distributed sensor and camera networks and biometric and forensic systems. Dr. Schmid was an organizer and the special session chair for the session on Theory of Biometric Systems for the 2008 IEEE International Conference on Acoustic, Speech, and Signal Processing. She has also served as a guest editor for the Special Issue on ``Recent Advances in Biometric Systems: A Signal Processing Perspective,'' published by the EURASIP Journal on Advances in Signal Processing. She is currently serving as a member of Editorial Board of International Scholarly Research Network (ISRN) Machine Vision, a member of Editorial Board of Dataset Papers in Physics, and a member of organizing committees for several conferences on human identification and image processing.
\end{IEEEbiography}
\vspace{-1 cm}
  \begin{IEEEbiography}[{\includegraphics[width=1in,height=1.25in,clip,keepaspectratio]{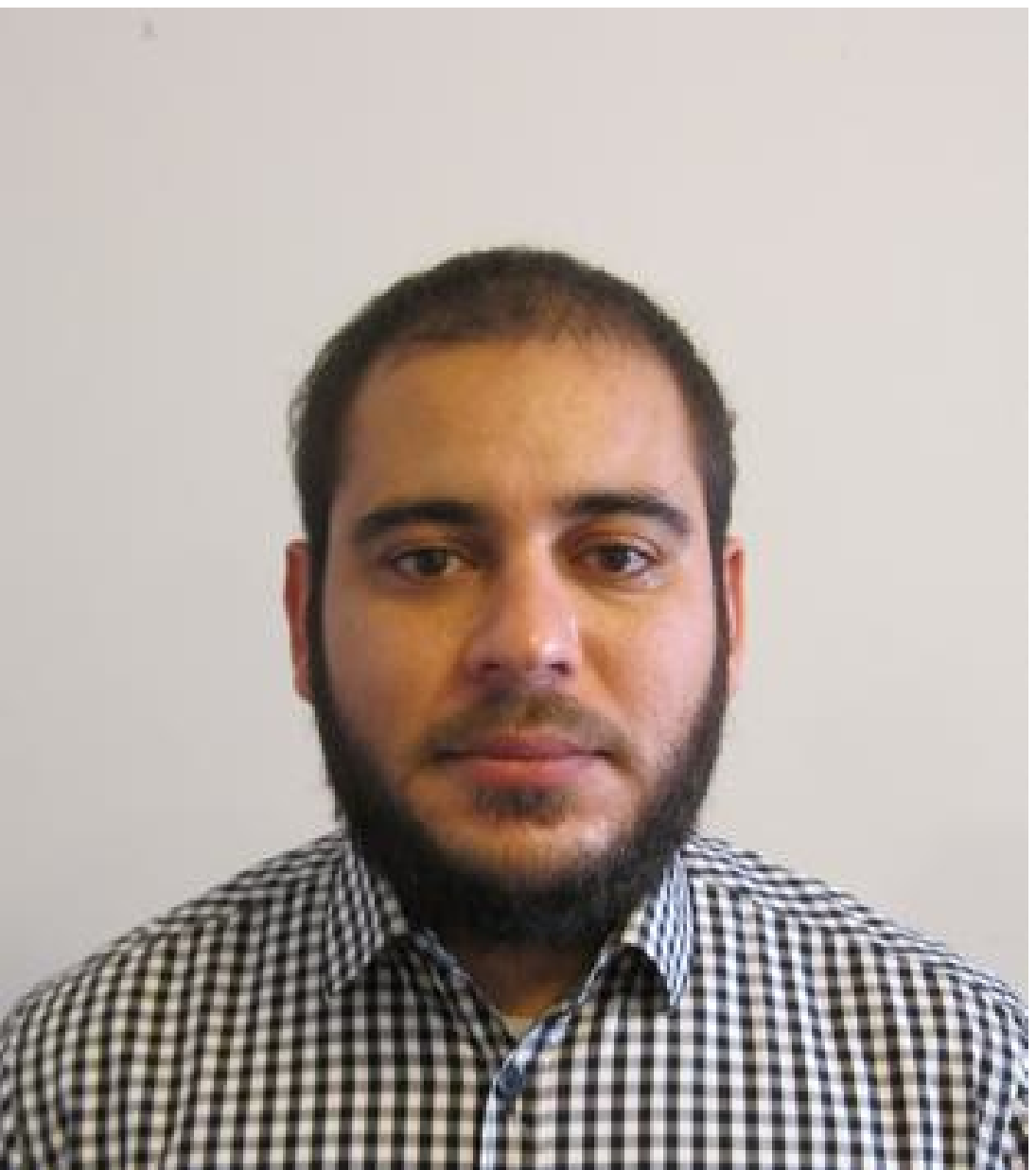}}]{Marwan Alkhweldi}
received the B.S. degree in Electrical Engineering from University of Tripoli, Libya, in 2007 and the M.S. degree in Electrical Engineering from West Virginia University, Morgantown, WV in 2012. He is currently pursuing the Ph.D. degree in Electrical Engineering at West Virginia University. His research interests are in the area of statistical signal processing and its applications to sensor networks and wireless communications.
\end{IEEEbiography}
\vspace{-1 cm}
  \begin{IEEEbiography}[{\includegraphics[width=1in,height=1.25in,clip,keepaspectratio]{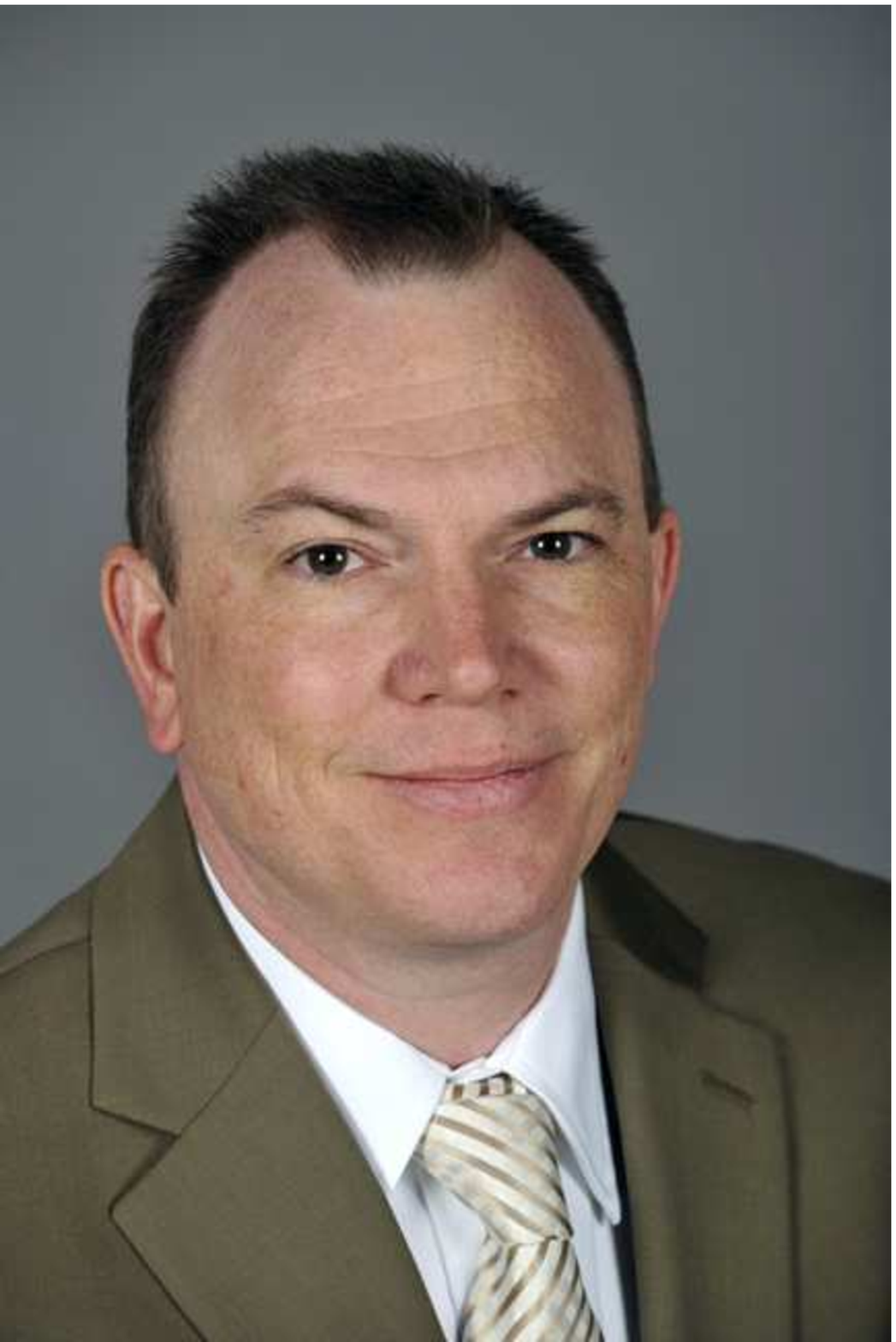}}]{Matthew C. Valenti}
is a Professor in Lane Department of Computer Science and Electrical Engineering at West Virginia University. He holds BS and Ph.D. degrees in Electrical Engineering from Virginia Tech and a MS in Electrical Engineering from the Johns Hopkins University. From 1992 to 1995 he was an electronics engineer at the US Naval Research Laboratory. He serves as an editor for {\em IEEE Wireless Communications Letters} and for {\em IEEE Transactions on Communications}.  He was Vice Chair of the Technical Program Committee for Globecom-2013.  Previously, he has served as a track or symposium co-chair for VTC-Fall-2007, ICC-2009, Milcom-2010, ICC-2011, and Milcom-2012, and has served as an editor for {\em IEEE Transactions on Wireless Communications} and {\em IEEE Transactions on Vehicular Technology}. His research interests are in the areas of communication theory, error correction coding, applied information theory, wireless networks, simulation, and secure high-performance computing.  His research is funded by the NSF and DoD.  He is registered as a Professional Engineer in the State of West Virginia.
\end{IEEEbiography}

\balance

\end{document}